\documentclass[journal,10pt]{IEEEtran}

\makeatletter
\def\ps@headings{%
	\def\@oddhead{\mbox{}\scriptsize\rightmark \hfil \thepage}%
	\def\@evenhead{\scriptsize\thepage \hfil \leftmark\mbox{}}%
	\def\@oddfoot{}%
	\def\@evenfoot{}}
\makeatother 

\usepackage[T1]{fontenc}
\usepackage{graphicx}
\usepackage{amsmath, amsfonts,epsfig, multirow, floatflt,float}
\usepackage{amssymb}
\usepackage{diagbox}
\usepackage{cite}
\usepackage{algorithm}
\usepackage{algorithmic}

\usepackage{hyperref}
\usepackage{enumitem}
\usepackage{mathrsfs}
\usepackage{url}
\usepackage{color}
\usepackage{subfigure}
\usepackage{array}
\usepackage{makecell}
\usepackage{tabularx}
\allowdisplaybreaks

\graphicspath{{figure/}}

\begin{document}
	
\title{Autonomous Platoon Control with Integrated Deep Reinforcement Learning and Dynamic Programming}
\author{Tong~Liu, Lei~Lei,  Kan~Zheng, Kuan~Zhang}

\maketitle

\begin{abstract}
Autonomous vehicles in a platoon determine the control inputs based on the system state information collected and shared by the Internet of Things (IoT) devices. Deep Reinforcement Learning (DRL) is regarded as a potential method for car-following control and has been mostly studied to support a single following vehicle. However, it is more challenging to learn an efficient car-following policy with convergence stability when there are multiple following vehicles in a platoon, especially with unpredictable leading vehicle behavior. In this context, we adopt an integrated DRL and Dynamic Programming (DP) approach to learn autonomous platoon control policies, which embeds the Deep Deterministic Policy Gradient (DDPG) algorithm into a finite-horizon value iteration framework. Although the DP framework can improve the stability and performance of DDPG, it has the limitations of lower sampling and training efficiency. In this paper, we propose an algorithm, namely Finite-Horizon-DDPG with Sweeping through reduced state space using Stationary approximation (FH-DDPG-SS), which uses three key ideas to overcome the above limitations, i.e., transferring network weights backward in time, stationary policy approximation for earlier time steps, and sweeping through reduced state space. In order to verify the effectiveness of FH-DDPG-SS, simulation using real driving data is performed, where the performance of FH-DDPG-SS is compared with those of the benchmark algorithms. Finally,  platoon safety and string stability for FH-DDPG-SS are demonstrated. \par
\end{abstract}

\begin{IEEEkeywords}
Platoon Control; Deep Reinforcement Learning; Dynamic Programming
\end{IEEEkeywords}
	
\section{Introduction}
As a key application scenario of Internet of Vehicles (IoV) \cite{Silva2019}, vehicle platooning has progressed dramatically thanks to the development of Cooperative Adaptive Cruise Control (CACC) and advanced Internet of Things (IoT) technologies in various fields \cite{Chen2020,Lesch2021,Ma2021,Lei2022,Huang2022,Zhang2022}. Autonomous platoon control aims to determine the control inputs for the following autonomous vehicles so that all the vehicles move at the same speed while maintaining the desired distances between each pair of preceding and following vehicles. The autonomous vehicles are considered to have L$3$ to L$5$ driving automation as defined by the Society of Automotive Engineers (SAE) classification standard \cite{SAE}. A well-designed platoon controller is able to increase road capacity, reduce fuel consumption, as well as enhance driving safety and comfort \cite{Li2017,Guo2020}. \par
	
Platoon controllers have been proposed based on classical control theory, such as linear controller, $\mathcal{H}_{\infty}$ controller, and Sliding Mode Controller (SMC) \cite{Li2017,Sinan2014,Yang2021}. Platoon control is essentially a Sequential Stochastic Decision Problem (SSDP), where a sequence of decisions have to be made over a specific time horizon for a dynamic system whose states evolve in the face of uncertainty. To solve such an SSDP problem, a few existing works relied on the Model Predictive Control (MPC) method \cite{Guo2020,Lan2020,Lin2020,MasseraFilho2017,vanNunen2019,Zheng2017}, where the trajectories of the leading vehicles are predicted by a model. \par 

Although the MPC controller provides some safety guarantees to the control policy, the control performance is still restricted by the accuracy of the model itself. As another promising method, Reinforcement Learning (RL) can learn an optimal control policy directly from experience data by trial and error without requiring the stochastic properties of the underlying SSDP model \cite{sutton2018reinforcement,o2018uncertainty,Heess2015a}. Moreover, the more powerful Deep Reinforcement Learning (DRL) methods can deal with the curse-of-dimensionality problem of RL by approximating the value functions as well as policy functions using deep neural networks \cite{hinton2006,Lei2019}. In recent years, research on DRL has made significant progress and many popular DRL algorithms have been proposed, including value-based methods such as Deep Q Network (DQN) \cite{Mnih2015} and Double DQN \cite{VanHasselt2016}; and actor-critic methods such as Deep Deterministic Policy Gradient (DDPG) \cite{lillicrap2015}, Asynchronous Advantage Actor-Critic (A3C) \cite{mnih2016}, and Trust Region Policy Optimization (TRPO) \cite{schulman2015}. The RL/DRL algorithms have been applied to solve the platoon control problem in a few recent literature\cite{Desjardins2011,Wang2014,Huang2019,Buechel2018,Li2018,wei2018design,Meixin2020,Lin2019,Lin2021,Zhang2020}. \par 

 To elaborate, most of the contributions have addressed the car-following control problem of supporting a single following vehicle. 
%	In \cite{Desjardins2011}, DRL is used for a CACC-based car-following control problem. 
An adaptive Proportional-Integral (PI) controller is presented in~\cite{Wang2014} whose parameters are tuned based on the state of the vehicle according to the control policy learned by actor-critic with kernel machines. However, the PI controller needs to predefine the candidate set of parameters before learning. In order to avoid this problem, an adaptive controller with parameterized batch actor-critic is proposed in~\cite{Huang2019}. 
%	The controller generates the exact throttle and brake commands instead of tuning the fuel/brake signals in the controller to track different speeds under different conditions.
 A few works improve the RL/DRL-based control policy by modelling/predicting the leading vehicle (leader)'s behavior\cite{Buechel2018,Li2018}. In \cite{Buechel2018}, a predictive controller based on the classical DRL algorithm DDPG \cite{lillicrap2015} is presented as an alternative to the MPC controller, which uses advanced information about future speed reference values and road grade changes. 
%	In addition, with the aid of a hybrid Markov process modeling the lead vehicle's speed, a drift-mitigation oriented optimal control-based approximate Q-learning algorithm is developed for Adaptive Cruise Control (ACC) systems in \cite{Li2018}. 
The human driving data has been used in \cite{wei2018design,Meixin2020} to help RL/DRL achieve improved performance. 
%	To improve training efficiency, a supervised RL method is proposed to improve the success rate of the training process, where the actor and critic networks are updated under the guidance of the supervisor and the gain scheduler \cite{wei2018design}. A velocity control scheme based on DDPG is proposed for car-following, which utilizes human driving data and driving features such as safety, efficiency, and comfort when designing the reward function \cite{Meixin2020}. 
In \cite{Lin2019}, the DDPG is applied for car-following control problem taking into account the acceleration-related delay, where the leader is assumed to drive at a constant speed. The proposed algorithm is used for comparing the performance of DRL and MPC for Adaptive Cruise Control (ACC)-based car-following control problems in \cite{Lin2021}. It is shown that DDPG has more advantages over MPC in the presence of uncertainties. In \cite{Zhang2020}, a deterministic promotion RL method is proposed to improve training efficiency, where the direction of action exploration is evaluated by a normalization-based function and the evaluated direction works as a model-free search guide in return.

Meanwhile, DRL-based platoon control with multiple following vehicles has only been studied in a few recent works \cite{wang2018novel,Chu2019,Yan2021}. Based on Predecessor-Leader Following (PLF) topology, a CACC-based control algorithm using DDPG is proposed in \cite{wang2018novel}. \textbf{While DDPG is the most widely used algorithm in the existing DRL-based car-following controllers \cite{Buechel2018, Meixin2020,Lin2019,Lin2021,wang2018novel}, it is shown that although DDPG performs well in the single following vehicle system, it is more difficult to learn an efficient control policy with convergence stability} in a platoon system where there are multiple following vehicles and unpredictable leading vehicle behavior \cite{Chu2019,Yan2021}. To address this problem, the DDPG-based technique is invoked in \cite{Chu2019} for determining the parameters of Optimal Velocity Model (OVM) instead of directly determining the accelerations. Meanwhile, \cite{Yan2021} proposes a Hybrid Car-Following Strategy (HCFS) that selects the best actions derived from the DDPG controller and the linear controller, which is used to determine vehicle acceleration in the platoon. \textbf{By combining with the classical control solutions, the performances are improved in \cite{Chu2019,Yan2021}. However, the classical controllers also limit the performance of the above solutions, especially in the complex driving environment with random disturbance and non-linear system dynamics.} \par

In this context, \textbf{we adopt an integrated DRL and Dynamic Programming (DP) approach to improve the convergence stability and performance of DDPG-based platoon control policy without resorting to the help of the classical controllers.} Specifically, we propose an algorithm that builds upon the Finite-Horizon DDPG (FH-DDPG) algorithm that was applied for the energy management of microgrids \cite{Lei2020}. FH-DDPG addresses the unstable training problem of DDPG in a finite-horizon setting by using two key ideas: backward induction and time-dependent actors/critics. The DDPG algorithm is embedded into a finite-horizon value iteration framework, and a pair of actor and critic networks are trained for each time step by backward induction. It has been demonstrated in \cite{Lei2020} that compared with DDPG, FH-DDPG is much more stable and achieves better performance.\par

However, FH-DDPG also suffers from some limitations that can be considered as the ``side-effects'' of its DP framework, i.e., low sampling efficiency and training efficiency. Firstly, since FH-DDPG has to train $K$ actor and critic networks for a finite-horizon problem with $K$ time steps, the sampling efficiency of FH-DDPG is $1/K$ that of DDPG. Specifically, for $E$ episodes of training experience, the actor and critic networks of DDPG are trained with $EK$ data entries, while each pair of the $K$ actor and critic networks in FH-DDPG is only trained with $E$ data entries at the corresponding time step. Secondly, FH-DDPG has to sweep through the entire state space when training the actor and critic networks at each time step. As a result, the exhaustive sweeps approach considers a large portion of the inconsequential states, resulting in many wasted training updates.\par

\textbf{To address the above two limitations in FH-DDPG and improve the sampling and training efficiency}, we use three key ideas in our proposed DRL algorithm for platoon control, namely FH-DDPG with Sweeping through reduced state space using Stationary policy approximation (FH-DDPG-SS). The contributions of this paper are itemized next and an approach summary of related works on DRL-based platoon controller design is given in Table \ref{table_summary} to describe characteristics of existing methods and highlight the contributions of our proposed algorithm.
\begin{table*} [!h]
    \centering
    \renewcommand{\arraystretch}{1.6}
    \setlength{\extrarowheight}{1pt}
    \setlength\tabcolsep{3pt} 
    \caption{Approach summary of related works on RL/DRL-based platoon controller design}
    \label{table_summary}
    {    
            \begin{tabularx}{\textwidth}{|c|c|c|X|}
            \hline
            \textbf{Scenario}&\textbf{Approach} & \textbf{References}&\makecell[c]{\textbf{General description}}\\
            \hline
            Single following vehicle &RL/DRL & \cite{Wang2014, Huang2019, Li2018,wei2018design, Zhang2020,Buechel2018, Meixin2020, Lin2019, Lin2021} & \multirow{2}{0.615\textwidth}{DDPG is the most widely used DRL algorithm and performs well in the single following vehicle system. \textbf{However}, it is more difficult to learn an efficient control policy with convergence stability in a platoon system where there are multiple following vehicles and unpredictable leading vehicle behavior.}   \\[1.2ex]
            \cline{1-2} \cline{3-3}
            \multirow{4}{*}{Multiple following vehicles}  &DDPG&\cite{wang2018novel}&     \\[1.2ex]
             \cline{2-4}
            &DDPG-OVM &\cite{Chu2019}&\multirow{2}{0.62\textwidth}{By combining with the classical control solutions, the performance of DDPG is improved. \textbf{However}, the classical controllers also limit the performance of these solutions, especially in the complex driving environment with random disturbance and non-linear system dynamics }\\[1.2ex]
              \cline{2-3}
            &HCFS&\cite{Yan2021} &  \\[1.2ex]
        \cline{2-4}
            &FH-DDPG-SS& Proposed& We adopt an integrated DRL and DP approach to \textbf{improve the convergence stability} and performance of DDPG-based platoon control policy without resorting to the help of the classical controllers. We also use three key ideas to overcome the limitations of the DP framework and \textbf{improve the sampling efficiency and training efficiency}.\\[2.2ex]
            \hline
           \end{tabularx}%
    }
\end{table*}
\begin{itemize}
	
%\item To overcome the first limitation of FH-DDPG, i.e., the low sampling efficiency, we propose two key ideas, namely transferring network weights backward in time and stationary policy approximation for earlier time steps. The first key idea is inspired by the parameter-transfer approach in transfer learning \cite{Pan2010}, where we transfer the trained actor and critic network weights at time step $k+1$ to the initial network weights at time step $k$. Thus, the actor and critic networks at time step $k$ are actually trained from the experiences of the $E(K-k)$ data entries of the $E$ episodes from time steps $k$ to $K$, which improves sampling efficiency. The second key idea is based on the observation that the optimal policies are approximately stationary for earlier time steps in a finite-horizon setting. Therefore, the time step threshold $m$ is first determined so the optimal policies and action-values are approximately stationary and constant from time steps $1$ to $m$. Next, we use FH-DDPG to train the actors and critics from time steps $K$ to $m+1$, and then train a single pair of actor and critic networks using DDPG from time steps $1$ to $m$, where the initial target network weights are set to the trained actor and critic network weights at time step $m+1$. The sampling efficiency is improved since the actor and critic networks are trained from the experiences of the $Em$ data entries from time steps $1$ to $m$. 

\item To overcome the first limitation of FH-DDPG, i.e., the low sampling efficiency, we propose two key ideas, namely\textbf{ transferring network weights backward in time} and \textbf{stationary policy approximation for earlier time steps}. The first key idea is to transfer the trained actor and critic network weights at time step $k+1$ to the initial network weights at time step $k$. The second key idea is using FH-DDPG to train the actors and critics from time steps $K$ to $m+1$, and then train a single pair of actor and critic networks using DDPG from time steps $1$ to $m$, where the initial target network weights are set to the trained actor and critic network weights at time step $m+1$. 

%\item  To address the second limitation of FH-DDPG, i.e., the wasteful updates due to exhaustive sweeps, we propose the third key idea, namely sweeping through reduced state space. A good "kick-off" policy is first learned by exhaustive sweeps, and then a reduced state space is obtained by testing the "kick-off" policy. Finally, the "kick-off" policy is trained by sweeping through the reduced state space to further improve the performance. This approach can help agents focus learning on the states that good policies visit often, so as to improve training efficiency.

\item  To address the second limitation of FH-DDPG, i.e., the wasteful updates due to exhaustive sweeps, we propose the third key idea, namely \textbf{sweeping through reduced state space.} Specifically, we train and test a ``kick-off'' policy to obtain a more refined state space. By sweeping through the reduced state space, the training efficiency and performance are enhanced as agents can focus on learning the states that good policies frequently visit.

\item To implement the above three key ideas, the FH-DDPG-SS algorithm is proposed to combine and integrate the three improvements for FH-DDPG.

\end{itemize} 

The remainder of the paper is organized as follows. The system model is introduced in Section II. Section III formulates the SSDP model for platoon control. The proposed DRL algorithms to solve the SSDP model are presented in Section IV. In Section V, the performance of FH-DDPG-SS is compared with those of the benchmark algorithms by simulation. Moreover, platoon safety and string stability are demonstrated. Section VI concludes the paper. 
	
\section{System Model}
We consider a platoon control problem with a number of $N>2$ vehicles, i.e., $\mathcal{V}=\{0,1,\cdots,N-1\}$, wherein the position, velocity and acceleration of a following vehicle (follower) $i\in \mathcal{V}\backslash \{0\}$ at time $t$ are denoted by $p_{i}(t)$, $v_{i}(t)$, and $acc_{i}(t)$, respectively. Here $p_{i}(t)$ represents the one-dimensional position of the center of the front bumper of vehicle $i$. Each follower $i$ is manipulated by a distributed car-following policy of a DRL controller with Vehicle-to-Everything (V2X) communications. \par

Each vehicle $i\in\mathcal{V}$ obeys the dynamics model described by a first-order system.
\begin{equation}
\label{eq2}
\dot{p}_{i}(t)=v_{i}(t),
\end{equation}
\begin{equation}
\label{eq3}
\dot{v}_{i}(t)=acc_{i}(t),
\end{equation}
\begin{equation}
\label{eq4}
\dot{acc}_{i}(t)=-\frac{1}{\tau_{i}}acc_{i}(t)+\frac{1}{\tau_{i}}u_{i}(t),
\end{equation}
\noindent where $\tau_{i}$ is a time constant representing driveline dynamics and $u_{i}(t)$ is the vehicle control input (commanded acceleration) at time $t$. In order to ensure driving safety and comfort, the following constraints are applied
\begin{equation}
\label{eq43}
acc_{\mathrm{min}} \leq acc_{i}(t)\leq acc_{\mathrm{max}},
\end{equation}
\begin{equation}
\label{eq44}
u_{\mathrm{min}} \leq u_{i}(t)\leq u_{\mathrm{max}},
\end{equation}
\noindent where $acc_{\mathrm{min}}$ and $acc_{\mathrm{max}}$ are the acceleration limits, while $u_{\mathrm{min}}$ and 	$u_{\mathrm{max}}$ are the control input limits. \par

The headway of follower $i$ at time $t$, i.e., bumper-to-bumper distance between follower $i$ and its preceding vehicle (predecessor) $i-1$, is denoted by $d_{i}(t)$ with
\begin{equation}
\label{eq5}
d_{i}(t)=p_{i-1}(t)-p_{i}(t)-L_{i-1},
\end{equation}
\noindent where $L_{i-1}$ is the the body length of vehicle $i-1$.

We adopt Constant Time-Headway Policy (CTHP) in this paper, i.e., follower $i$ aims to maintain a desired headway $d_{r,i}(t)$, which satisfies
\begin{equation}
\label{eq6}
d_{r,i}(t)=r_{i}+h_{i}v_{i}(t),
\end{equation}
\noindent where $r_{i}$ is a standstill distance for safety of follower $i$ and $h_{i}$ is a constant time gap of follower $i$ which represents the time that it takes for follower $i$ to bridge the distance in between the vehicles $i$ and $i-1$ when continuing to drive with a constant velocity.

The control errors, i.e., gap-keeping error $e_{pi}(t)$ and velocity error $e_{vi}(t)$ of follower $i$ are defined as
\begin{equation}
\label{eq7}
e_{pi}(t)=d_{i}(t)-d_{r,i}(t),
\end{equation}
\begin{equation}
\label{eq8}
e_{vi}(t)=v_{i-1}(t)-v_{i}(t).
\end{equation}

\section{SSDP Model for Platoon Control}

An SSDP can be formulated to determine the vehicle's control action. The time horizon is discretized into time intervals of length $T$ seconds (s), and a time interval $[(k-1)T,kT)$ amounts to a time step $k$, $k = 1,2,\cdots,K$, where $K$ is the total number of time steps. In the rest of the paper, we will use $x_{k}:=x((k-1)T)$ to represent any variable $x$ at time $(k-1)T$.

In the following, the state space, action space, system dynamics model, and reward function of the SSDP
model are presented, respectively.\par
	
\subsection{State Space}	
At each time step $k\in \{1,2,\cdots, K\}$, the controller of follower $i$ determines the vehicle control input $u_{i,k}$, based on the observations of the system state. $v_{i,k}$ and $acc_{i,k}$ can be measured locally, while $e_{pi,k}$ and $e_{vi,k}$ can be measured through a radar unit mounted at the front of the vehicle. Thus, the state that the follower $i$ can obtain locally is denoted by $x_{i,k}=[e_{pi,k},e_{vi,k},acc_{i,k}]^{\mathrm{T}}$.\par

Additionally, the follower $i$ can obtain the driving status $x_{j,k}$ and control input $u_{j,k}$ of the other vehicles $j\in\mathcal{V}\backslash \{i\}$ via V2X communications. In this paper, we adopt the Predecessors Following (PF) information topology, in which $acc_{i-1,k}$ and $u_{i-1,k}$ are transmitted to the follower $i\in\mathcal{V}\backslash \{0\}$. It has been proved in our previous work that the optimal policy of the SSDP model with the preceding vehicle's acceleration $acc_{i-1,k}$ and control input $u_{i-1,k}$ in the state performs at least as well as the optimal policy for the SSDP model which does not include this information \cite{Lei2022}. Thus, the system state for the follower $i$ is denoted as: $S_{i,k}=[x_{i,k},acc_{i-1,k},u_{i-1,k}]^{\mathrm{T}} $. The state space is $\mathcal{S}=\{S_{i,k}|e_{pi,k},e_{vi,k} \in[-\infty,\infty],acc_{i,k},acc_{i-1,k}\in[acc_{\mathrm{min}},acc_{\mathrm{max}}],u_{i-1,k}\in[u_{\mathrm{min}},u_{\mathrm{max}}]\}$.

%$=\bigcup_{i=1}^{N-1}[e_{p,\mathrm{min}},e_{p,\mathrm{max}}]\times[e_{v,\mathrm{min}},e_{v,\mathrm{max}}]\times[acc_{\mathrm{min}},acc_{\mathrm{max}}]\times[acc_{\mathrm{min}},acc_{\mathrm{max}}]$, where $e_{p,\mathrm{min}}$ and $e_{p,\mathrm{max}}$ are gap-keeping error limits, $e_{v,\mathrm{min}}$ and $e_{v,\mathrm{max}}$ are velocity error limits. \par
	
% 	\begin{figure}[htb!]
% 		\centering
% 		\includegraphics[width=0.51\textwidth]{IFT.pdf}
% 		\caption{V2X information transmitted to vehicle $i$ for a platooning scenario.}
% 		\label{fig_1}
% 	\end{figure}
 
\subsection{Action Space}	
The control input $u_{i,k}$ of the follower $i\in\mathcal{V}\backslash \{0\}$ is regarded as the action at the time step $k$\footnote{An action is normally denoted as $a$ in the RL literature. In this paper, we adopt the convention in the optimal control literature and denote the action as $u$ for consistency.}. The action space is $\mathcal{A}=\{u_{i,k}|u_{i,k}\in [u_{\mathrm{min}},u_{\mathrm{max}}]\}$.

\subsection{System Dynamics Model}		
The system dynamics are derived in discrete time on the basis of forward Euler discretization. Note that for the leader $0$, $e_{p0,k}=e_{v0,k}=0$, thus the system dynamics model evolves in discrete time according to
	\begin{equation}
	\label{eq17}
	x_{0,k+1}=A_{0}x_{0,k}+B_{0}u_{0,k},
	\end{equation}
	\noindent where
	\begin{equation}
	\label{eq11}
	A_{0}=\begin{bmatrix}
	0 & 0 & 0 \\
	0 & 0 & 0 \\
	0 & 0 & 1-\frac{T}{\tau_{0}}
	\end{bmatrix},
	B_{0}=\begin{bmatrix}
	0\\
	0 \\
	\frac{T}{\tau_{0}}
	\end{bmatrix}.	
	\end{equation}	
 	For the follower $i\in\mathcal{V}\backslash \{0\}$ in the platoon,  we have
 	
	\begin{equation}
	\label{eq12}
	x_{i,k+1}=A_{i}x_{i,k}+B_{i}u_{i,k}+C_{i}acc_{i-1,k},
	\end{equation}
	\noindent where
	\begin{equation}
	\label{eq13}
	A_{i}=\begin{bmatrix}
	1 & T & -h_{i}T \\
	0 & 1 & -T \\
	0 & 0 & 1-\frac{T}{\tau_{i}}
	\end{bmatrix},
	B_{i}=\begin{bmatrix}
	0\\
	0 \\
	\frac{T}{\tau_{i}}
	\end{bmatrix},
	C_{i}=\begin{bmatrix}
	0\\
	T \\
	0
	\end{bmatrix}.		
	\end{equation}		
 \par
	
\subsection{Reward Function}	

%	The platoon's control goal is to ensure plant- and string-stability so that a desired headway can be achieved. It is plant stable when disturbances in velocity are attenuated along the entire platoon and string stable when a platoon approaches the constant velocity of its leader. \cite{Chu2019}.\par
Reward function can guide the optimization objectives and has an impact on the convergence of the DRL algorithm. Our objective is to minimize gap-keeping error $e_{pi,k}$ and velocity error $e_{vi,k}$ to achieve the platoon control target while penalizing control input $u_{i,k}$ to reduce the fuel consumption and the jerk to improve the driving comfort. Note that the jerk is the change rate in acceleration, which is given by 
\begin {align}	
	j_{i,k}&=\frac {acc_{i,k+1}-acc_{i,k}}{T}=-\frac{1}{\tau_{i}}acc_{i,k}+\frac{1}{\tau_{i}}u_{i,k}, 
\end{align}
\noindent where the second equality is due to the forward Euler discretization of \eqref{eq4}.\par

Although the quadratic cost function is normally adopted in optimal control problems, it is found that it does not work well for DDPG algorithm as the sudden large changes of reward values often decrease its training stability. Therefore, an absolute-value cost function is adopted for DDPG in \cite{Lin2019,Lin2021,Yan2021} to improve its performance. However, we found that the absolute-value cost function could hinder the further performance improvement when the control errors are relatively small. Therefore, we design a Huber loss function \cite{huber2004} as the reward function for each follower $i\in \mathcal{V}\backslash \{0\}$, which is given by 
	
\begin{equation}
	\label{eq14}
	R(S_{i,k},u_{i,k})=\left\{
	\begin{array}{ll}
		r_{\mathrm{abs}}, & \mathrm{if} \ r_{\mathrm{abs}}< \varepsilon\\
		r_{\mathrm{qua}}, & \mathrm{if} \ r_{\mathrm{abs}}\geq \varepsilon \\
	\end{array}\right. ,
\end{equation}

	\noindent where
	\begin{equation}
	\label{eqabs}
	 r_{\mathrm{abs}}=-\{|\frac{e_{pi,k}}{\hat{e}_{p,\mathrm{max}}}|+a|\frac{e_{vi,k}}{\hat{e}_{v,\mathrm{max}}}|+b|\frac{u_{i,k}}{u_{\mathrm{max}}}|+c|\frac{j_{i,k}}{2acc_{\mathrm{max}}/T}|\},\IEEEnonumber
	\end{equation}
	 \begin{equation}
	 r_{\mathrm{qua}}=-\lambda \{(e_{pi,k})^2+a(e_{vi,k})^2+b(u_{i,k})^2+c(j_{i,k}T)^2\},\IEEEnonumber
	\end{equation}
	  where $\varepsilon$ is the reward threshold, $\hat{e}_{p,\mathrm{max}}$ and $\hat{e}_{v,\mathrm{max}}$ are the nominal maximum control errors such that  it is larger than most possible control errors. $\lambda$ is the reward scale. $a$, $b$ and $c$ are the positive weights and can be adjusted to determine the relative importance of minimizing the gap-keeping error, the velocity error, the control input and the jerk. \par
	Thus, the expected cumulative reward $J_{\pi_i}$ of the follower $i$ over the finite time horizon $K$ under a policy $\pi_i$ can be expressed as
	\begin{equation}
            \label{eq18}
		J_{\pi_i}=\mathrm{E}_{\pi_i}[\sum_{k=1}^{K} \gamma^{k-1} R(S_{i,k},u_{i,k})],
	\end{equation}
where $\gamma$ is the reward discount factor.

	The ultimate objective is to find the optimal policy $\pi_i^{*}$ that maximizes the expected cumulative reward $J_{\pi_i}$, i.e.,
	\begin{equation}
		\label{eq19}
		\pi_i^{*}=\arg\max_{\pi_i}J_{\pi_i}.
	\end{equation}   
	
%	Additionally, the convergence of the DRL algorithm is often affected by the designed reward function. For DDPG algorithm, the sudden large changes of reward values often decrease its training stability. Therefore, to better show the car following the performance of DDPG for comparison with the algorithm we proposed, we use the normalized reward function.

\section{DRL Algorithms}
To solve the above SSDP, we propose a DRL algorithm which improves on the FH-DDPG algorithm\cite{Lei2020}. In the following, we will first provide a brief introduction to the FH-DDPG algorithm, and then elaborate on the proposed improvements.
\subsection{FH-DDPG}
FH-DDPG is a combination of DRL and DP, where the DDPG algorithm is embedded into a finite-horizon value iteration framework. It is designed to solve finite-horizon SSDP and improve the stability of the DDPG algorithm. \par

DDPG is a well-known DRL algorithm widely applied to continuous control. It trains both a pair of actor and critic networks, i.e., $\mu(S_k|\theta^\mu)$ and $Q(S_k, u_k |\theta^Q)$, to derive the optimal policy $\mu^*(s|\theta^{\mu})$ and the corresponding action-value (Q-value) $Q^*(S_k,u_k|\theta^{Q})$, respectively\cite{lillicrap2015}. The action-value is defined as the expected cumulative discounted reward from time step $k$: $Q(S_k, u_k|\theta^Q)=\mathrm{E}_{\pi}[\sum_{k'=k}^{\infty} \gamma^{k'-k} R(S_{k'},u_{k'})]$. Experience replay is adopted in DDPG to enable stable and robust learning. When the replay buffer is full, the oldest sample will be discarded before a new sample is stored in the buffer. A minibatch from the buffer is sampled at each time step in order to update the actor and critic networks. DDPG creates a copy of the actor and critic networks as target networks, i.e., $\mu'(S_{k+1}|\theta^{\mu'})$ and $Q'(S_{k+1},u_{k+1}|\theta^{Q'})$, to calculate the target values. The critic network is updated by minimizing the Root Mean Square Error (RMSE) $L_k = R(S_k,u_k)+\gamma Q'(S_{k + 1},\mu'(S_{k + 1}|\theta^{\mu'})|\theta^{Q'})-Q(S_k, u_k|\theta^Q)$ using the sampled gradient descent with respect to $\theta^Q$. The actor network is updated by using the sampled deterministic policy gradient ascent on $Q(S_{k },\mu(S_{k }|\theta^{\mu})|\theta^{Q})$ with respect to $\theta^\mu$. In order to improve the stability of learning, the weights of these target networks are updated by soft target update, where the target networks are constrained to slowly track the learned networks. 

\begin{figure}[htpb!]
	\centering
	\includegraphics[width=0.43\textwidth]{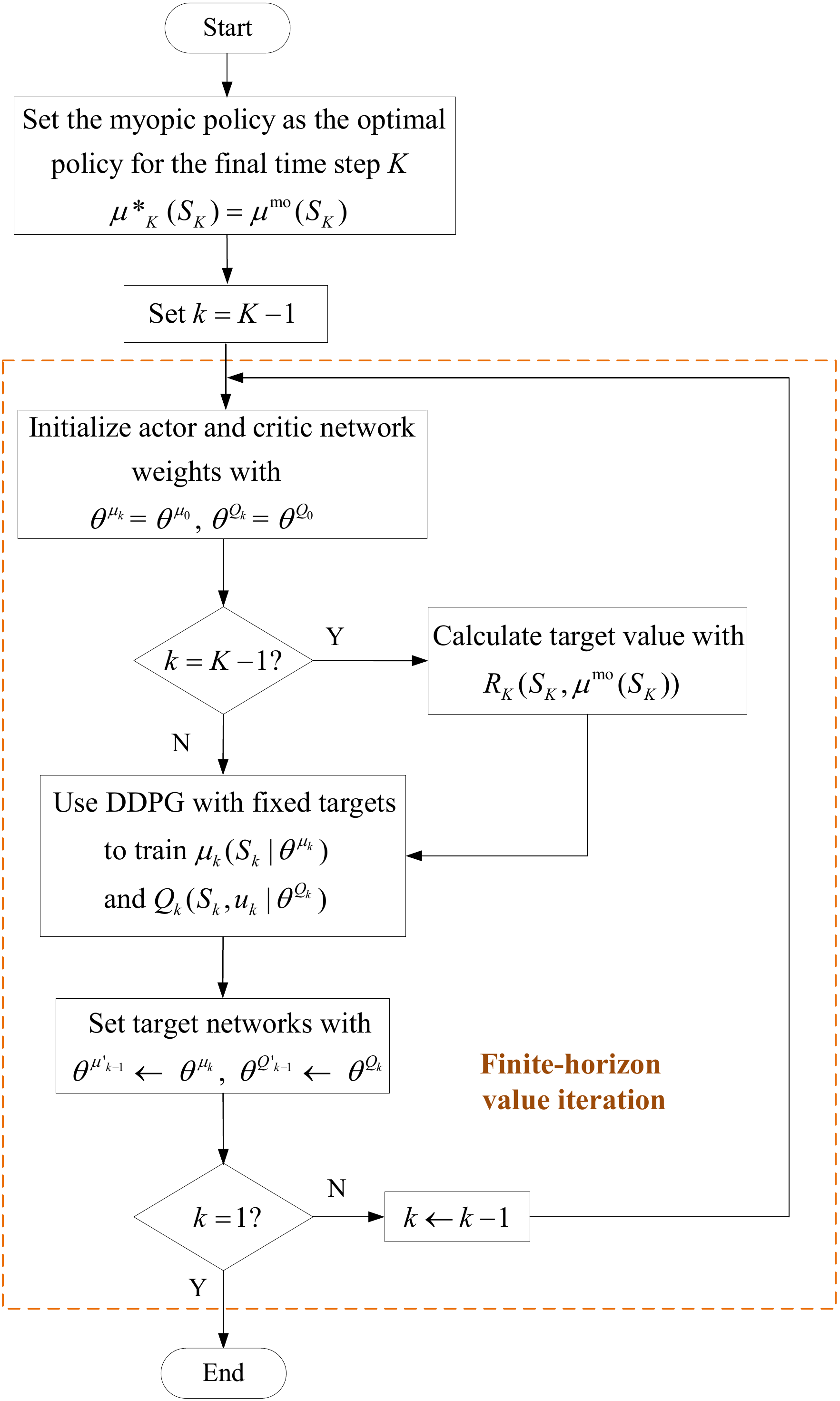}
	\caption{FH-DDPG framework.}
	\label{fig_1}
\end{figure}

DDPG is designed to solve the infinite-horizion SSDPs, where the actors and critics are the same for every time step. On the other hand, the optimal policies and the corresponding action-values are normally time-dependent in a finite-horizon setting \cite{puterman2014}. Therefore, there are $K$ actors and critics in FH-DDPG for an SSDP with $K$ time steps. As shown in Fig.\ref{fig_1}, FH-DDPG starts by having the myopic policy $\mu_K^*(S_K)=\mu^{\mathrm{mo}}(S_K)$ as the optimal policy with the terminal reward $R_K$ for the final time step $K$. And then, the finite horizon value iteration starts from time step $K-1$, and uses backward induction to iteratively derive the optimal policy $\mu^*_{k}(S_k|\theta^{\mu_k})$ and the action-value $Q^*_{k}(S_k,u_k|\theta^{Q_k})$ for each time step $k$, until it reaches the first time step $k=1$. In each time step, an algorithm similar to DDPG is adopted to solve a one-period MDP in which an episode only consists of two time steps. However, different from DDPG, the target actor network $\mu'_{k}(S_{k+1}|\theta^{\mu'_k})$ and critic network $Q'_{k}(S_{k+1},u_{k+1}|\theta^{Q'_k})$ of the current time step $k$ are fixed and set as the trained actor network $\mu_{k+1}(S_{k+1}|\theta^{\mu_{k+1}})$ and critic network $Q_{k+1}(S_{k+1},u_{k+1}|\theta^{Q_{k+1}})$ of the next time step $k+1$. This greatly increases the stability and performance of the algorithm. The pseudocode of the FH-DDPG algorithm is given in Appendix A. Note that in each time step, the $\mathrm{DDPG-FT}$ function is used to train the respective actor and critic networks, where $\mathrm{DDPG-FT}$ is the abbreviation for DDPG with fixed targets.\par

\subsection{Improving sampling efficiency}
%Compared with DDPG that is trained with $EK$ data entries for $E$ episodes, the actor and critic networks per time step in FH-DDPG are only trained with $E$ data entries. The sample efficiency of FH-DDPG is $1/K$ that of DDPG. To improve sampling efficiency, two improvements are proposed in the following.

For a finite-time horizon consisting of $K$ time steps, when both DDPG and FH-DDPG are trained for $E$ episodes, a total of $EK$ data entries are generated from the experience data as a data entry is generated per time step. Since DDPG has a single pair of actor and critic networks, all the $EK$ data entries are used to train this pair of networks. Meanwhile, there are $K$ pairs of actor and critic networks in FH-DDPG. For each time step $k$, the corresponding pair of actor and critic networks is trained only with the $E$ data entries of time step $k$. Therefore, the sampling efficiency of FH-DDPG is only $1/K$ of that of DDPG. To improve sampling efficiency, two key ideas are proposed in the following.
\subsubsection{Transferring network weights backward in time}
In FH-DDPG, the actor ${\mu}_{i,k}(S_{i,k}| \theta^{\mu_{i,k}})$ and the critic ${Q}_{i,k}(S_{i,k},u_{i,k}|\theta^{Q_{i,k}})$ at each time step $k$ are trained with random initial parameters. Inspired by the parameter-transfer approach in transfer learning\cite{Pan2010}, we transfer the trained actor and critic network weights at time step $k+1$, i.e., $\theta^{\mu_{i,k+1}}$ and $\theta^{Q_{i,k+1}}$ to the initial network weights at time step $k$, i.e., $\theta^{\mu_{i,k}}$ and $\theta^{Q_{i,k}}$ respectively. Thus, although $\mu_{i,k}(S_{i,k}|\theta^{\mu_{i,k}})$ and $Q_{i,k}(S_{i,k},u_{i,k}|\theta^{Q_{i,k}})$ are trained based on the $E$ data entries of the $E$ episodes at time step $k$, the trainings are built upon the initial weights $\theta^{\mu_{i,k+1}}$ and $\theta^{Q_{i,k+1}}$, which were in turn trained based on the $E$ data entries of the $E$ episodes at time step $k+1$ and built upon the initial weights $\theta^{\mu_{i,k+2}}$ and $\theta^{Q_{i,k+2}}$. In this way, the actor and critic networks at time step $k$ are actually trained from the experiences of the $E(K-k)$ data entries of the $E$ episodes from time steps $k$ to $K$, instead of only the $E$ data entries as in FH-DDPG. The proposed algorithm is given in Algorithm~\ref{alg2}, namely FH-DDPG with Network weights transferring Backward in time (FH-DDPG-NB).   \par
 \begin{algorithm}
 	\renewcommand{\algorithmicrequire}{\textbf{Input:}}
 	\renewcommand{\algorithmicensure}{\textbf{Output:}}
	\caption{FH-DDPG-NB algorithm}
	\label{alg2}
	\begin{algorithmic}[1]

	\STATE Randomly initialize actor and critic network weights as $\theta^{\mu0}$ and $\theta^{Q0}$	
	\STATE Set $\mu_{i,K}^*(S_{i,K})=\mu^{\mathrm{mo}}(S_{i,K})$ for the final time step $K$
	\FOR{$k= K - 1, \cdots, 1$}
		\IF{$k=K-1$}
			\STATE $\theta^{\mu_i}=\theta^{\mu0}$ and $\theta^{Q_i}=\theta^{Q0}$ 
		\ELSE
			\STATE $\theta^{\mu_{i}}=\theta^{\mu_{i,k+1}}$ and $\theta^{Q_{i}} =\theta^{Q_{i,k+1}}$
		\ENDIF
			\STATE 	$\theta^{\mu_{i,k}},\theta^{Q_{i,k}}   \leftarrow \mathrm{DDPG-FT}(\theta^{\mu_{i}},\theta^{Q_{i}},\theta^{\mu'_i},\theta^{Q'_i},k)$
	\STATE Update the target network:
	\begin{displaymath}
	\theta^{\mu'_i}\leftarrow\theta^{\mu_{i,k}},\ \theta^{Q'_i}\leftarrow\theta^{Q_{i,k}}
	\end{displaymath}
	\ENDFOR
	\STATE \textbf{return} $\{\theta^{\mu_{i,k}},\theta^{Q_{i,k}}\}_{k=1}^{K-1} $
	
\end{algorithmic}
\end{algorithm}

\subsubsection{Stationary policy approximation for earlier time steps} 
Although the action-values are generally time-dependent for the finite-horizon control problems, there is a vanishing difference in action-values when the horizon length is sufficiently large \cite{vossen2022finite}. Taking the finite-horizon Linear Quadratic Regulator (LQR) as an example, the action-value gradually converges to the steady-state value as the time step $k$ decreases from the horizon $K$. Moreover, for $k$ not close to horizon $K$, the LQR optimal policy is approximately stationary \cite{LQR2009,LQR2018}. Even though the platoon control problem is not an LQR problem in the strict sense, since both the system state $S_{i,k}$ and action $u_{i,k}$ have constraints and there are non-Gaussian random disturbances, we can observe similar trends in the learned policy by FH-DDPG. This allows us to improve sampling efficiency by first obtaining the time step threshold $m$ such that the action-values and optimal policies are approximately constant and stationary when $k\leq m$. And then, we adopt a single pair of actor and critic networks from time steps $1$ to $m$. 

To elaborate, we first obtain $m$ by solving the LQR problem for platoon control and analyzing the corresponding results, ignoring the state/action constraints and random disturbances. This enables us to determine the value of $m$ in an efficient manner. Then FH-DDPG is trained from time steps $K$ to $m+1$. Next, instead of training a separate pair of actor and critic networks for each time step from $1$ to $m$, we train a single actor network $\mu_{i}(S_{i,k}|\theta^{\mu_{i}})$ and critic network $Q_{i}(S_{i,k}, u_{i,k}|\theta^{Q_{i}})$ for all the time steps $k\in\{1,2,...,m\}$. Specifically, the actor and critic networks are trained using DDPG, where the initial values of the target networks are set to those of the trained actor and critic networks at time step $m+1$, i.e., $\theta^{\mu_{i,m+1}}$ and $\theta^{Q_{i,m+1}}$. The well-trained initial values for the target networks can significantly increase the stability and performance of the DDPG algorithm. In this way, the actor and critic networks are trained from the experiences of the $Em$ data entries of the $E$ episodes from time steps $1$ to $m$. The proposed algorithm is given in Algorithm~\ref{alg3}, namely FH-DDPG with Stationary policy Approximation for earlier time steps (FH-DDPG-SA). The function $\mathrm{FH-DDPG}$ is realized by the FH-DDPG algorithm given in Appendix A. Note that FH-DDPG-SA can be combined with FH-DDPG-NB, by adopting the latter algorithm instead of FH-DDPG to train the actor and critic networks from time steps $K$ to $m+1$  in line \ref{alg2_2} of the pseudocode. This will result in the FH-DDPG-SA-NB algorithm. 
%
%\textbf{Therefore, a "kick-off" policy $\hat{\mu}_{i,k}(S_{i,k}|\hat \theta^{\mu_{i,k}})$ for the follower $i$ for time step $k$ is learned and then we obtain the upper and lower bounds of the more refined state space $\mathcal{S}_{i,k}^{\mathrm{rs}}$ for the follower $i$ at time step $k$ by testing $\hat \mu_{i}(S_{i,k}|\hat \theta^{\mu_{i}}),k=1,2,\dots,m$
%and $\hat \mu_{i,k}(S_{i,k}|\hat \theta^{\mu_{i,k}}),k=m+1,m+2,\dots,K-1$. Next, the actor network $\hat{\mu}_{i,k}(S_{i,k}|\hat \theta^{\mu_{i,k}})$ and critic network $\hat{Q}_{i,k}(S_{i,k},u_{i,k}|\hat\theta^{Q_{i,k}})$ are further trained by FH-DDPG and DDPG, which only sweeps through $\mathcal{S}_{i,k}^{\mathrm{rs}}$.} 

  \begin{algorithm}
  	\renewcommand{\algorithmicrequire}{\textbf{Input:}}
  	\renewcommand{\algorithmicensure}{\textbf{Output:}}
 	\caption{FH-DDPG-SA-(NB) algorithm}
 	\label{alg3}
 	\begin{algorithmic}[1]	
		\STATE Set the time horizon as $\{m+1,\cdots,K\}$ 
		\STATE $ \{\theta^{\mu_{i,k}}, \theta^{Q_{i,k}}\}_{k=m+1}^{K-1}   \leftarrow \mathrm{FH-DDPG(-NB)}$ \label{alg2_2}
		\STATE Set the time horizon as $\{1,\cdots,m\}$ 
		\STATE 	Set the initial target networks weights with $\theta^{\mu'_{i}} = \theta^{\mu_{i,m+1}}$ and $ \theta^{Q'_{i}} = \theta^{Q_{i,m+1}}$
		\STATE $ \theta^{\mu_{i}}, \theta^{Q_{i}}  \leftarrow \mathrm{DDPG}$
 	\end{algorithmic}
 \end{algorithm}

\subsection{Sweeping through reduced state space}	
FH-DDPG embeds DRL under the framework of DP, while the classical approach of DP is to sweep through the entire state space at each time step $k$. This exhaustive sweeps approach leads to many wasteful updates during training, since many of the states are inconsequential as they are visited only under poor policies or
with very low probability. An alternative approach is trajectory sampling which sweeps according to on-policy distribution\cite{sutton2018reinforcement}. Although trajectory sampling is more appealing, it is impossible to be adopted by FH-DDPG due to the latter's backward induction framework.

Inspired by trajectory sampling, we improve FH-DDPG by sweeping through a reduced state space. Specifically, we first learn a relatively good ``kick-off'' policy by exhaustive sweeps, and then obtain a reduced state space by testing the ``kick-off'' policy, and finally continue to train the policy by sweeping through the reduced state space to further improve the performance. This approach can help agents to focus on learning the states that good policies frequently visit, which improves training efficiency. Note that a good policy in RL should achieve a large expected cumulative reward $J_{\pi_i}$ as given in (18). For example in platoon control, the control errors are normally small under good policies as the ends of the training episodes are approached, and it is not necessary to sweep through large control error states.

For platoon control, although the theoretical bounds of gap-keeping error $e_{pi,k}$ and velocity error $e_{vi,k}$ are infinity, it is impossible to sweep through an infinite range when training. Therefore, we need to restrict sweeping to a finite range at first. In practice, there are some empirical limits to $e_{pi,k}$ and $e_{vi,k}$ for a reasonable platoon control policy. Since we consider that FH-DDPG is trained from scratch, some relatively large control error states could be visited during training due to the random initial policy and exploration. Therefore, we first sweep through a relatively large state space, i.e.,
\begin{align}
\label{19}
\mathcal{S}^{\mathrm{ls}}=&\{[e_{pi,k},e_{vi,k},acc_{i,k}]^{\mathrm{T}}|\IEEEnonumber\\
&e_{pi,k}\in[-e_{p,\mathrm{max}},e_{p,\mathrm{max}}],\IEEEnonumber\\
&e_{vi,k}\in[-e_{v,\mathrm{max}},e_{v,\mathrm{max}}],\IEEEnonumber\\
&acc_{i,k}\in[acc_{\mathrm{min}},acc_{\mathrm{max}}]\},
\end{align}
where $e_{p,max}$ and $e_{v,max}$ are the same for each time step and are larger than most control errors during the training of FH-DDPG. Thus, we first train FH-DDPG in the state space $\mathcal{S}^{\mathrm{ls}}$ to learn a ``kick-off'' policy $\hat{\mu}_{i,k}(S_{i,k}|\hat \theta^{\mu_{i,k}})$ for the follower $i$ at time step $k$, and then obtain the upper and lower bounds of a more refined state space, i.e.,  
\begin{align}
\label{20}
\mathcal{S}_{i,k}^{\mathrm{rs}}=&\{[e_{pi,k},e_{vi,k},acc_{i,k}]^{\mathrm{T}}|\IEEEnonumber\\
&e_{pi,k} \in[\overline e_{pi,k,\mathrm{min}},\overline e_{pi,k,\mathrm{max}}],\IEEEnonumber\\
&e_{vi,k}\in[\overline e_{vi,k,\mathrm{min}},\overline e_{vi,k,\mathrm{max}}],\IEEEnonumber\\
&acc_{i,k}\in[\overline {acc}_{i,k,\mathrm{min}},\overline {acc}_{i,k,\mathrm{max}}]\},
\end{align}
for the follower $i$ at time step $k$ by testing $\hat{\mu}_{i,k}(S_{i,k}|\hat \theta^{\mu_{i,k}})$. Next, the actor network $\hat{\mu}_{i,k}(S_{i,k}|\hat \theta^{\mu_{i,k}})$ and critic network $\hat{Q}_{i,k}(S_{i,k},u_{i,k}|\hat\theta^{Q_{i,k}})$ are further trained by FH-DDPG, which only sweeps through $\mathcal{S}_{i,k}^{\mathrm{rs}}$. 

Combining the above three improvements for FH-DDPG, we propose a novel DRL algorithm, namely FH-DDPG with Sweeping through reduced state space using Stationary policy approximation (FH-DDPG-SS), which is given in Algorithm~\ref{alg1}. Note that the overall procedure of FH-DDPG-SS is the same as that described in Section IV.C, except that in line \ref{alg3_2} and line \ref{alg3_12}, the FH-DDPG-SA-NB and FH-DDPG-SA algorithms are adopted instead of the FH-DDPG algorithm to incorporate the improvements in Algorithm~\ref{alg2} and Algorithm~\ref{alg3}. The reason why we use FH-DDPG-SA instead of FH-DDPG-SA-NB in line \ref{alg3_12} is that the initial actor and critic networks weights for all the time steps are carried over from the previous training, so we no longer need to transfer network weights backward in time. \par

%	 where $\mathrm{FH-DDPG}(\theta^{\mu_{0}},\theta^{Q_{0}},\mathcal{S}^{\mathrm{ls}})$ and  $\mathrm{FH-DDPG}(\hat\theta^{\mu_{i,k}},\hat\theta^{Q_{i,k}},\mathcal{S}_{i,k}^{\mathrm{rs}})$ are obtain by function $\mathrm{FH-DDPG}(\widetilde \theta^{\mu_{i,k}},\widetilde\theta^{Q_{i,k}},\widetilde {\mathcal{S}}_{i,k})$ given in Appendix A}

\begin{algorithm}
	\caption{FH-DDPG-SS algorithm}
	\label{alg1}
	\begin{algorithmic}[1]	
		\STATE Set $\mathcal{S}^{\mathrm{ls}}$ according to ~\eqref{19}
		%			\STATE Randomly initialize actor and critic network weights as $\theta^{\mu_{0}}$ and $\theta^{Q_{0}}$
		\STATE $\{\hat\theta^{\mu_{i,k}},\hat\theta^{Q_{i,k}}\}_{k=1}^{K-1} \leftarrow \mathrm{FH-DDPG-SA-NB}(\mathcal{S}^{\mathrm{ls}})$ \label{alg3_2}
		\FOR {$g=1,\dots, G$}
		\STATE Test $ \{\hat \mu_{i,k}(S_{i,k}|\hat \theta^{\mu_{i,k}})\}_{k=1}^{K-1}$ 
		\STATE Store $\{x_{i,k}^{(g)}\}_{k=1}^{K-1} =\{[e_{pi,k}^{(g)},e_{vi,k}^{(g)}, acc_{i,k}^{(g)}]^{\mathrm{T}}\}_{k=1}^{K-1}$ in $\{\mathcal{B}_{i,k}^{\mathrm{t}}\}_{k=1}^{K-1}$
		\ENDFOR	
		\FOR {$k= 1, \dots, K-1$}
		%		the minimum and maximum value, i.e., 
		\STATE Find the upper and lower bounds $\overline e_{pi,k,\mathrm{min}}$, $\overline e_{pi,k,\mathrm{max}}$, $\overline e_{vi,k,\mathrm{min}}$, $\overline e_{vi,k,\mathrm{max}}$, $\overline {acc}_{i,k,\mathrm{min}}$ and $\overline {acc}_{i,k,\mathrm{max}}$ in $\mathcal{B}_{i,k}^{\mathrm{t}}$
		
		\STATE Set $\mathcal{S}_{i,k}^{\mathrm{rs}}$ according to \eqref{20}
			%			 the reduced state space $\mathcal{S}_{i,k}^{\mathrm{rs}}$ with stored  $S_{i,k}$ 
	
		\ENDFOR
		\STATE Set the initial actor and critic network weights as $\{\hat\theta^{\mu_{i,k}},\hat\theta^{Q_{i,k}}\}_{k=1}^{K-1}$
		\STATE $\{\theta^{\mu_{i,k}},\theta^{Q_{i,k}}\}_{k=1}^{K-1}   \leftarrow \mathrm{FH-DDPG-SA}(\{\mathcal{S}_{i,k}^{\mathrm{rs}}\}_{k=1}^{K-1})$ \label{alg3_12}	
	\end{algorithmic}
\end{algorithm}

\section{Experimental Results}		
	In this section, we present the simulation results of the proposed FH-DDPG-SS algorithm as well as benchmark DRL algorithms, i.e., DDPG, FH-DDPG, and HCFS \cite{Yan2021}. \par

\subsection{Experimental Setup}
	All the algorithms are trained/tested using the open source data given in \cite{Meixin2020}. Specifically, the data was extracted from the Next Generation Simulation (NGSIM) dataset \cite{NGSIM2009}, which was retrieved from the eastbound I-80 in the San Francisco Bay area in Emeryville, CA, on April $13$, $2005$, at a sampling rate of $10$ Hz, with $45$ minutes of precise location data available in the full dataset. Then, car-following events were extracted by applying a car-following filter as described in \cite{Wang2018}, where the leading and following vehicle pairs of each event stay in the same lane. In our experiment, we used $1000$ car-following events. Moreover, although there are data for both the leading vehicle and following vehicle in the dataset, we only used the velocity data of the leading vehicle to simulate the real-world environment with uncertainty, so that the DRL algorithms can be trained and evaluated. $80\%$ of the data (i.e., $800$ car-following events) is used for training and $20\%$ (i.e., $200$ car-following events) is used for testing. The platoon control environment and the DRL algorithms are implemented in Tensorflow 1.14 using Python~\cite{Liu2016}. We compare the performance of the proposed FH-DDPG-SS algorithm with the benchmark algorithms in terms of the average cumulative reward. Moreover, the platoon safety and string stability performance for FH-DDPG-SS are also demonstrated.
    To ensure a fair comparison, all the algorithms are trained and tested in the same environment with the same reward function.
	
	The technical constraints and operational parameters of the platoon control environment are given in Table~\ref{table_1}. In general, the parameters of the platoon environment and state/action space in Table I are determined using the values reported in \cite{Lin2019, Lin2021} as a reference. The interval for each time step is set to $T = 0.1$ $\rm s$, and each episode is comprised of $100$ time steps (i.e., $K=100$) with a duration of $10$ $\rm s$. As the number of vehicles set in the existing literature on platoon control ranges from $3$ to $8$\cite{Sinan2014,Lan2020,vanNunen2019,Zheng2017,wang2018novel,Chu2019,Yan2021}, we set the number of vehicles to $N=5$. We initialize the states for each of $4$ followers with $x_{i,0}=[1.5,-1,0]$, $\forall i\in\{1,2,3,4\}$. We experiment with several sets of reward coefficients $a, b$, and $c$ to balance the minimization of gap-keeping error $e_{pi,k}$, velocity error $e_{vi,k}$, control input $u_{i,k}$, and jerk $j_{i,k}$. In addition, the absolute-value cost function can improve the algorithm's convergence stability when control errors are large, and the square-value cost function is conducive to further performance improvement when control errors are small. Thus, we determine the reward threshold $\varepsilon$ with the best performance for all algorithms by a simple grid search. The nominal maximum control errors in the reward function \eqref{eq14} are set to $\hat{e}_{p,\rm max}=15$ $\rm m$ and $\hat{e}_{v,\rm max}=10$ $\rm m/s$ so that it is larger than most possible control errors during training for all DRL algorithms. For the FH-DDPG-SS algorithm, the time step threshold $m=11$. Moreover, the maximum gap-keeping error $e_{p,\rm max}$ and maximum velocity error $e_{v,\rm max}$ in \eqref{19} are set to $2$ $\rm{m}$ and $1.5$ $\rm{m/s}$, respectively. Additionally, to reduce large oscillations in $u_{i,k}$ and $acc_{i,k}$, we set $j_{i,k}$ in FH-DDPG and FH-DDPG-SS in the testing phase within $[-0.3,0.6]$ when $k>11$ by clipping $u_{i,k}$.
	
	\begin{table}[htb!]
			\renewcommand{\arraystretch}{1.2}
		\setlength\tabcolsep{1.8pt}  
		\centering
		\caption{Technical constraints and operational parameters of the platoon control environment}
		\begin{tabular}{lll}
			\hline
			\textbf{Notations} & \textbf{Description}&\textbf{Values} \\
			\hline
			\textbf{Platoon environment}&&\\
			
			$T$ &Interval for each time step & $0.1$ $\rm s$\\
                
			$K$&Total time steps in each episode  &$100$\\
			 $m$&     Time step threshold & 11\\
			$N$&Number of vehicles &5\\
			$\tau_i$&Driveline dynamics time constant  & $0.1$ $\rm s$\\
	
			$h_i$ &Time gap & $1$ s\\
			\hline
			\textbf{State \& action}&&\\
			$e_{p,\rm max}$&Maximum gap-keeping error  & $2$ $\rm{m}$\\
%			$e_{v,\rm min}$&Minimum velocity error  & $-1.5 \rm{m/s}$\\
			$e_{v,\rm max}$&Maximum velocity error  & $1.5$ $\rm{m/s}$\\
			$acc_{\rm min}$&Minimum acceleration & $-2.6$  $\rm{m/s^2}$\\
			$acc_{\rm max}$&Maximum acceleration   & $2.6$ $\rm{m/s^2}$\\
			$u_{\rm min}$&Minimum control input   & $-2.6$ $\rm{m/s^2}$\\
			$u_{\rm max}$&Maximum control input   & $2.6$ $\rm{m/s^2}$\\
			\hline
			\textbf{Reward function}&&\\
			$a$&Reward coefficient & $0.1$\\
			$b$ &Reward coefficient & $0.1$\\
			$c$ &Reward coefficient & $0.2$\\
			$	\hat{e}	_{p,\rm max}$&Nominal maximum gap-keeping error  & $15$ $\rm{m}$\\
			$	\hat{e}	_{v,\rm max}$&Nominal maximum velocity error  & $10$ $\rm{m/s}$\\
			$\varepsilon$&Reward threshold & $-0.4483$\\
			\hline
		\end{tabular}
		\label{table_1}
	\end{table}

	\begin{table*}[!h]
	\renewcommand{\arraystretch}{1.3}
	\caption{Hyper-Parameters of the DRL algorithms for training} \label{alg_para} \centering
	\begin{tabular}{llll}
		\hline
		{\textbf{Parameter}} & \multicolumn{3}{c} {\textbf{Value}} \\
		\hline
		& DDPG &FH-DDPG&FH-DDPG-SS\\
		\hline
		Actor network size &$256,128$ &  $400,300,100$& $400,300,100$\\
		\hline
		Critic network size & $256,128$ & $400,300,100$& $400,300,100$\\
		\hline
		Actor activation function & relu, relu, tanh & relu, relu, relu, tanh& relu, relu, relu, tanh\\ 
		\hline
		Critic activation function & relu, relu, linear& relu, relu, relu, linear & relu, relu, relu, linear\\ 
		\hline
		Actor learning rate $\alpha$ & $1$e$-4$ &$1$e$-4$&$1$e$-4$  \\
		\hline
		Critic learning rate $\beta$ & $1$e$-3$&$1$e$-3$ &$1$e$-3$ \\

		\hline
		Total training episodes $E$ & $5000$&$5000$ &$3000,2000$\\
		\hline
		Batch size $N_b$ & \multicolumn{3}{c} {$64$} \\
				\hline
		Replay buffer size & $250000$ &$2500$&$2500,2000$ \\
		\hline
		Reward scale $\lambda$ &  \multicolumn{3}{c} {$5$e$-3$} \\
		\hline
		Reward discount factor $\gamma$ &\multicolumn{3}{c} {1}\\
		\hline
		Soft target update $\eta$&$0.001$&  /    &  FH-DDPG: /, DDPG: $0.001$\\
		\hline
		Noise type & \multicolumn{3}{c}{Ornstein-Uhlenbeck Process with $\theta=0.15$ and $\sigma=0.5$} \\
		\hline
		Final layer weights/biases initialization & \multicolumn{3}{c}{Random uniform distribution $[-3\times10^{-3},3\times10^{-3}]$} \\
		\hline
		Other layer weights biases initialization &\multicolumn{3}{c}{\makecell[c]{Random uniform distribution$[-\frac{1}{\sqrt{f}},\frac{1}{\sqrt{f}}]$\\($f$ is the fan-in of the layer)}} \\
		\hline
		\label{table_2}
	\end{tabular}
\end{table*}
	
	 The hyper-parameters for training are summarized in Table \ref{table_2}. The values of all the hyper-parameters were selected by performing a grid search as in \cite{Mnih2015}, using the values reported in \cite{lillicrap2015} as a reference. DDPG has two hidden layers with $256$ and $128$ nodes, respectively; while FH-DDPG and FH-DDPG-SS have three hidden layers with $400$, $300$, and $100$ nodes, respectively. The sizes of input layers for all DRL algorithms are the same and decided by the PF information topology. Moreover, an additional $1$-dimensional action input is fed to the second hidden layer for each critic network. The total number of training episodes $E$ for all DRL algorithms is set to $5000$. For FH-DDPG-SS, we first train the algorithm for $3000$ episodes to learn the ``kick-off'' policy in the first phase, and then continue to train $2000$ episodes within the reduced state space in the second phase. The replay buffer sizes for DDPG and FH-DDPG are $250000$ and $2500$, respectively. This is because the replay buffer for FH-DDPG only stores the data entries for the corresponding time step. Since FH-DDPG-SS is trained in two phases, the replay buffer sizes for the first and second phases are $2500$ and $2000$, respectively. Moreover, FH-DDPG-SS leverages the FH-DDPG-SA-(NB) algorithm, which trains the $K-m-1$ actors and critics for time steps $K$ to $m+1$ using FH-DDPG, and a single pair of actor and critic for time steps $1$ to $m$ using DDPG. The soft target update is implemented with a parameter of $\eta=0.001$ for DDPG. As FH-DDPG uses a fixed target network, there is no soft target update.

\subsection{Comparison of FH-DDPG-SS with the benchmark algorithms}
\subsubsection{Performance for testing data}
The individual performance of each follower $i \in\{1,2,3,4\}$ as well as the sum performance of the $4$ followers are reported in Table~\ref{table_3} for DDPG, FH-DDPG, HCFS, and FH-DDPG-SS, respectively. The individual performance of each follower and the sum performance of all followers are obtained by averaging the returns of the corresponding followers and the sum returns of all followers, respectively, over $200$ test episodes after training is completed. Note that in RL terminology, a return is the cumulative rewards of one episode. It can be observed that the individual performance of the preceding vehicles are attenuated by following vehicles upstream the platoon for each algorithm. Compared with the other algorithms, FH-DDPG-SS consistently shows the best individual performance for each follower $i\in\{1,2,3,4\}$. We can observe that the ranking in terms of the sum performance of all followers for the different algorithms is FH-DDPG-SS$>$HCFS$>$FH-DDPG$>$DDPG, where FH-DDPG-SS outperforms DDPG, FH-DDPG, and HCFS algorithms by $46.62\%$, $13.73\%$, and $10.60\%$, respectively. Note that HCFS outperforms DDPG because two actions are obtained from DDPG and the classical linear controller at each time step, and the one that achieves the maximum reward is selected. On the other hand, FH-DDPG performs better than DDPG as it applies backward induction and time-dependent actors/critics. Moreover, FH-DDPG-SS further improves the performance of FH-DDPG by implementing the three key ideas proposed in Section IV.\par

In Table \ref{table_4}, we present the maximum, minimum, and standard deviation of the sum returns of the $4$ followers across the $200$ test episodes for DDPG, FH-DDPG, HCFS, and FH-DDPG-SS, respectively. It can be observed that although the objective of DRL algorithms is to maximize the expected return as given in \eqref{eq18} and \eqref{eq19}, FH-DDPG-SS achieves the best performance in terms of the maximum, minimum, and standard deviation of the returns among all the DRL algorithms.

\begin{table*}[htb!]
	\renewcommand{\arraystretch}{1}
	\setlength{\extrarowheight}{1pt}
	\centering
	\caption{Performance after training with NGSIM dataset. Each episode has $100$ time steps in total. We present the individual performance of each follower as well as the sum performance of the 4 followers for DDPG, FH-DDPG, HCFS, and FH-DDPG-SS, respectively.}
	\begin{tabular}{ |c|cccc|c|}
		\hline
		\multirow{2}{*}{\textbf{Algorithm}}&\multicolumn{4}{c|}{\textbf{ Individual performance}}&\multirow{2}{*}{\textbf{Sum performance}}\\
		\cline{2-5}
		&\textbf{Follower 1}&\textbf{Follower 2}&\textbf{Follower 3}&\textbf{Follower 4}&\\
		\hline
		DDPG&-0.0680&-0.0876&-0.0899&-0.2980&-0.5435\\ 
		\cline{1-6}
		FH-DDPG&-0.0736&-0.0845&-0.0856&-0.0927&-0.3364\\
		\cline{1-6}
		HCFS&-0.0673&-0.0740&-0.0828&-0.1005&-0.3246\\ 
		\cline{1-6}
		FH-DDPG-SS&-0.0600&-0.0691&-0.0776&-0.0835 &-0.2902\\
		\hline	
	\end{tabular}
	\label{table_3}
\end{table*}

\begin{table*}[htb!]
	\renewcommand{\arraystretch}{1}
	\setlength{\extrarowheight}{1pt}
	\centering
	\caption{The maximum, minimum, and standard deviation of the sum returns of the $4$ followers across the $200$ test episodes for DDPG, FH-DDPG, HCFS, and FH-DDPG-SS, respectively.}
	\begin{tabular}{|c|ccc|}
		\hline
		\textbf{Algorithm} & \textbf{Maximum}& \textbf{Minimum} &\textbf{Standard deviation}\\
		\hline
		DDPG&-0.5018&-0.5822&0.0137 \\
		\hline
		FH-DDPG&-0.3248&-0.3532&0.0071\\
		\hline
		HCFS&-0.3011&-0.3458&0.0068\\
		\hline
		FH-DDPG-SS&-0.2726&-0.3114&0.0063\\
		\hline
	\end{tabular}
	\label{table_4}
\end{table*}

\subsubsection{Convergence properties}

\begin{figure*}[!t]
	\centering
	\subfigure[DDPG]{
			\includegraphics[scale=0.235]{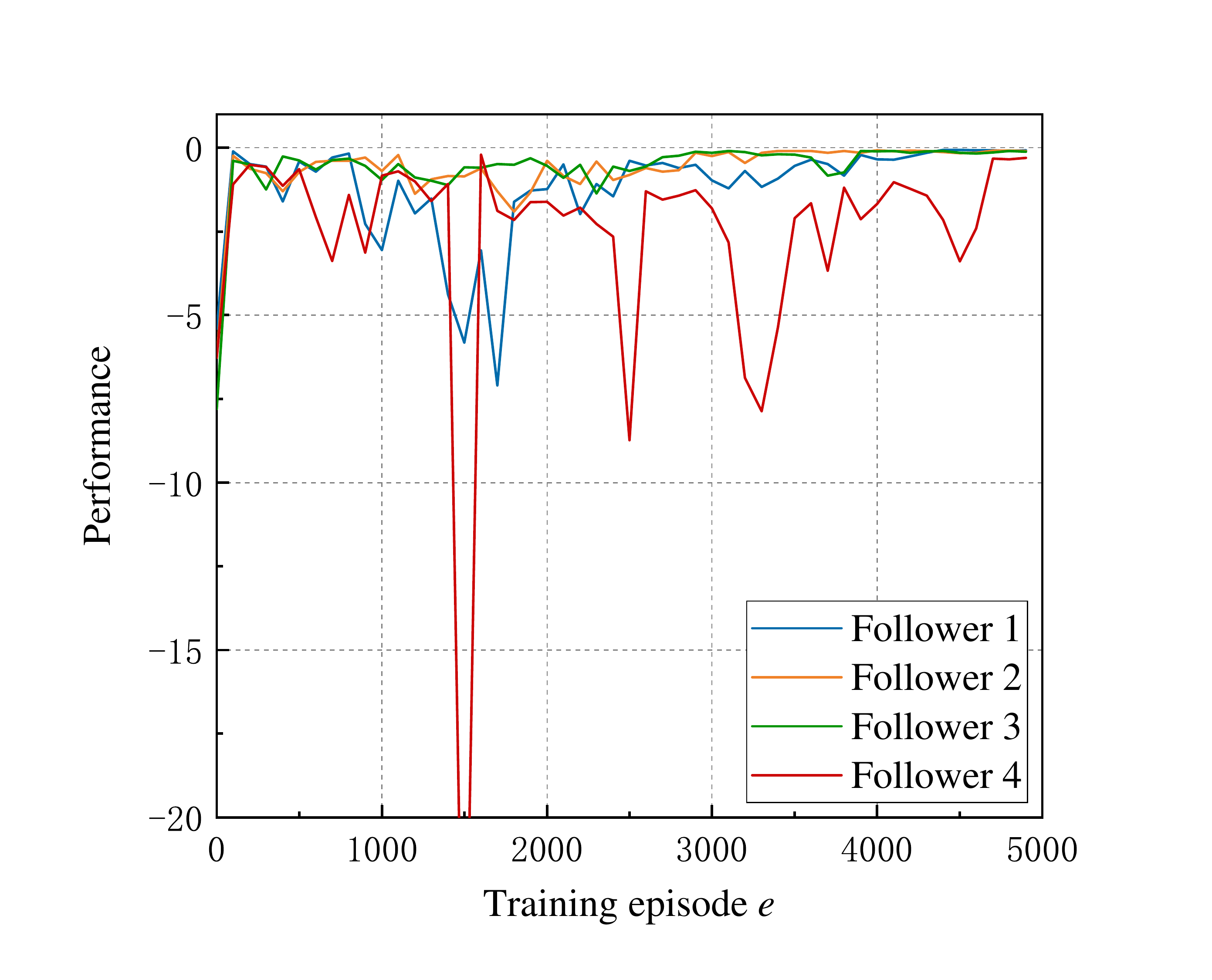}
		\label{fig2_sub1}
	}	
	\subfigure[FH-DDPG]{
			\includegraphics[scale=0.235]{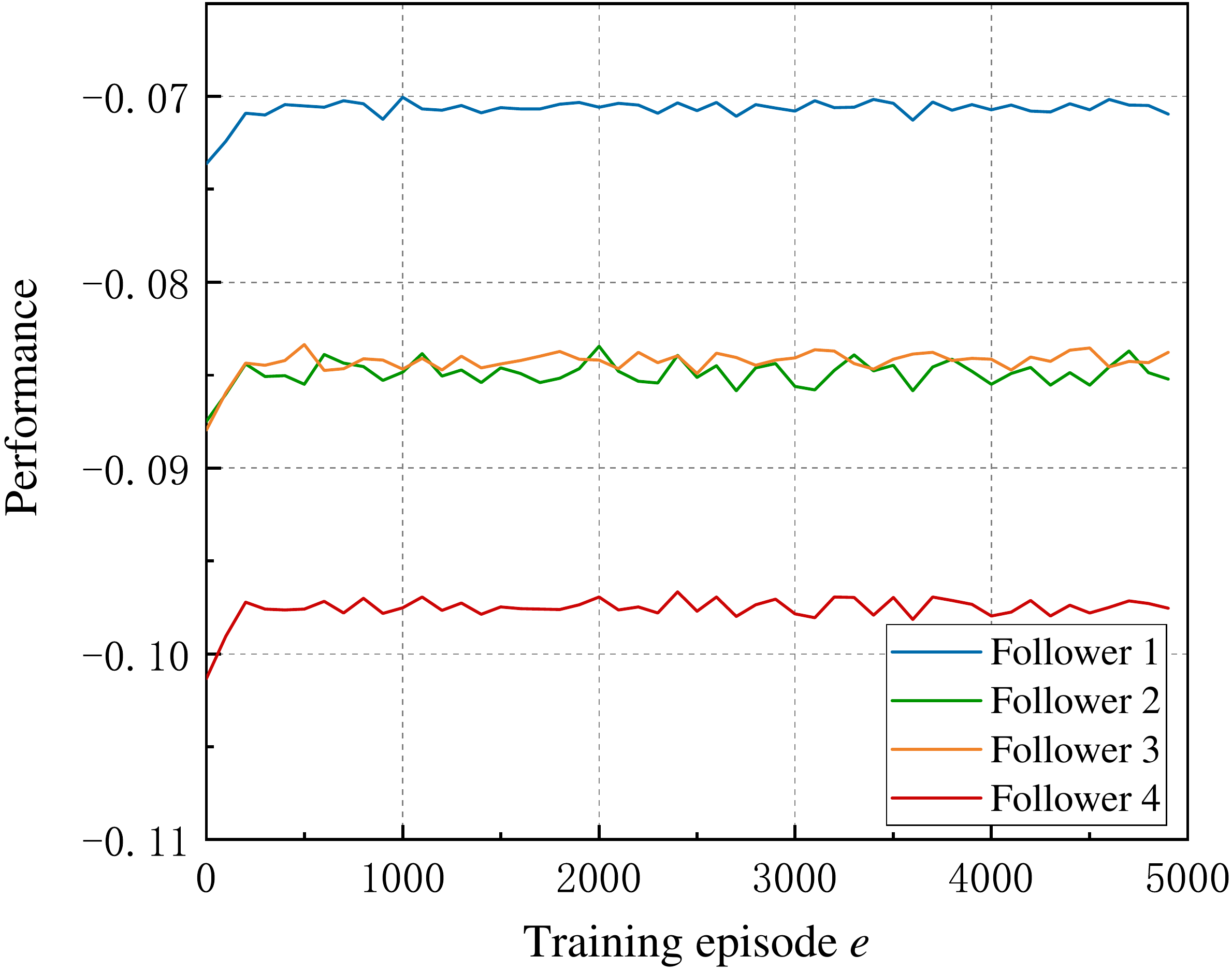}
		\label{fig2_sub2}
	} 
	\subfigure[FH-DDPG-SS]{
			\includegraphics[scale=0.235]{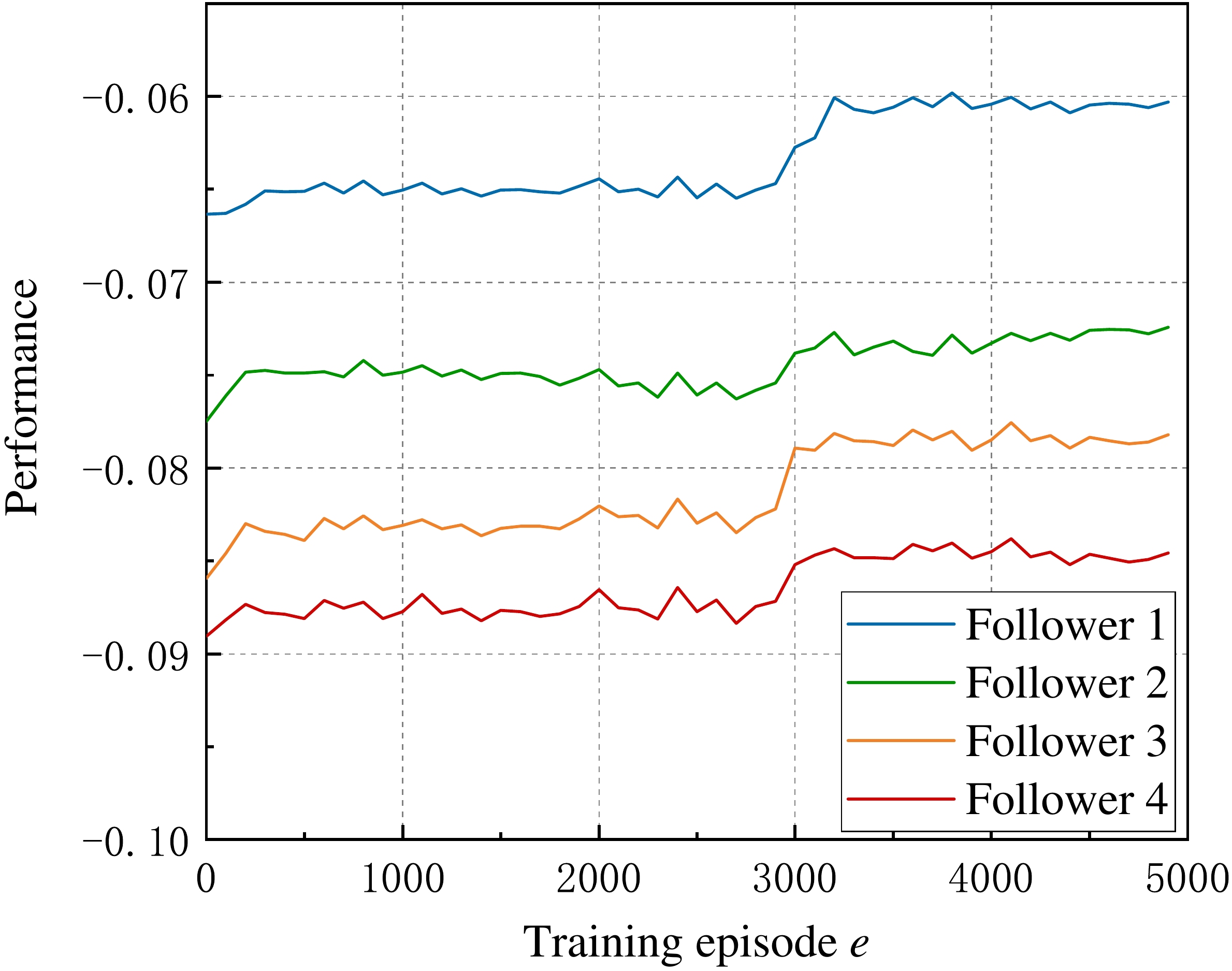}
		\label{fig2_sub3}
	}
	\caption{Performance during DRL algorithms training. The vertical axis corresponds to the average returns over 10 test episodes.}
	\label{fig_2}
\end{figure*}
The performance of DRL algorithms is evaluated periodically during training by testing without exploration noise. Specifically, we run $10$ test episodes after every $100$ training episodes, and average the returns over the $10$ test episodes as the performance for the latest $100$ training episodes. The performance as a function of the number of training episodes for each follower $i$ with DDPG, FH-DDPG, and FH-DDPG-SS is plotted in Fig.~\ref{fig_2}. The convergence curve of HCFS is not plotted here since HCFS combines the trained DDPG controller with the linear controller, and thus the convergence property of HCFS is the same as that of DDPG. It can be observed from Fig.~\ref{fig_2} that the initial performances of FH-DDPG and FH-DDPG-SS are much better than that of DDPG. This is due to the different ways how DDPG and FH-DDPG/FH-DDPG-SS work. For DDPG, the first point in Fig.~\ref{fig2_sub1} is obtained by running $10$ test episodes of the initial policy and calculating the average return. Since the initial weights of the actor network are random, the initial policy has poor performance. In FH-DDPG, there is a pair of actor/critic networks associated with each time step, and the actors/critics are trained backward in time. Therefore, when we train the actor/critic of the first time step, the actors/critics of the rest of the time steps are already trained. Similarly, the first point in Fig.~\ref{fig2_sub3} for FH-DDPG-SS is obtained by running $10$ test episodes where only the policies (actors) for time steps $1$ to $m$ are not trained,  while the policies (actors) of the time steps $m+1$ to $K$ are already trained.

Fig.~\ref{fig2_sub1} shows that although the performances of all the followers in DDPG improve quickly in the first $200$ episodes, the performance curves exhibit significantly larger oscillation during the following training episodes compared to those of FH-DDPG and FH-DDPG-SS. This is especially true for follower $4$, whose performance drops below $-20$ at around $1500$ episodes. The above results demonstrate that the convergence of DDPG is relatively unstable, especially for the followers upstream the platoon. As shown in Fig.~\ref{fig2_sub2}, the convergence stability of FH-DDPG is significantly improved over DDPG, as there are only small fluctuations in the performance curve. This is due to the fact that FH-DDPG trains backward in time, where DDPG with a fixed target is adopted in each time step to solve a one-period MDP. However, as FH-DDPG is under the framework of dynamic programming, it sacrifices the sampling and training efficiency to achieve the convergence stability. As shown in Fig.~\ref{fig2_sub2}, although the performance of FH-DDPG quickly increases in the first $200$ episodes, there is hardly any further improvement during the remaining training episodes. Fig.~\ref{fig2_sub3} shows that the convergence stability of FH-DDPG-SS is similar to that of FH-DDPG and much better than that of DDPG. Moreover, the performance of FH-DDPG-SS is consistently better than that of FH-DDPG for each follower during the whole training episode. For example, focusing on follower $1$, the performance of FH-DDPG is around $-0.0735$ at the beginning of training and finally reached $-0.0715$ towards the end. On the other hand, the performance of FH-DDPG-SS is around $-0.0665$ initially and finally reached $-0.0600$ at the end of training. The superior performance of FH-DDPG-SS before $3000$ episodes is due to the two proposed key ideas of transferring network weights backward in time and stationary policy approximation for earlier time steps, which improve the sampling efficiency. In other words, from the same number of data entries generated by the $3000$ training episodes, FH-DDPG-SS can make more efficient use of the data entries to learn a better policy. Moreover, it can be observed that there is a sudden performance improvement for all the vehicles of FH-DDPG-SS at around $3000$ episodes. This salient performance gain is due to the third key idea of sweeping through the reduced state space. Since FH-DDPG-SS is trained in the refined state space for the last $2000$ episodes, the training efficiency is improved as agents focus on learning the states that good policies frequently visit.

\subsubsection{Testing results of one episode}

\begin{figure*}[htbp!]
	\centering 
	\subfigure[DDPG]{
			\includegraphics[scale=0.3]{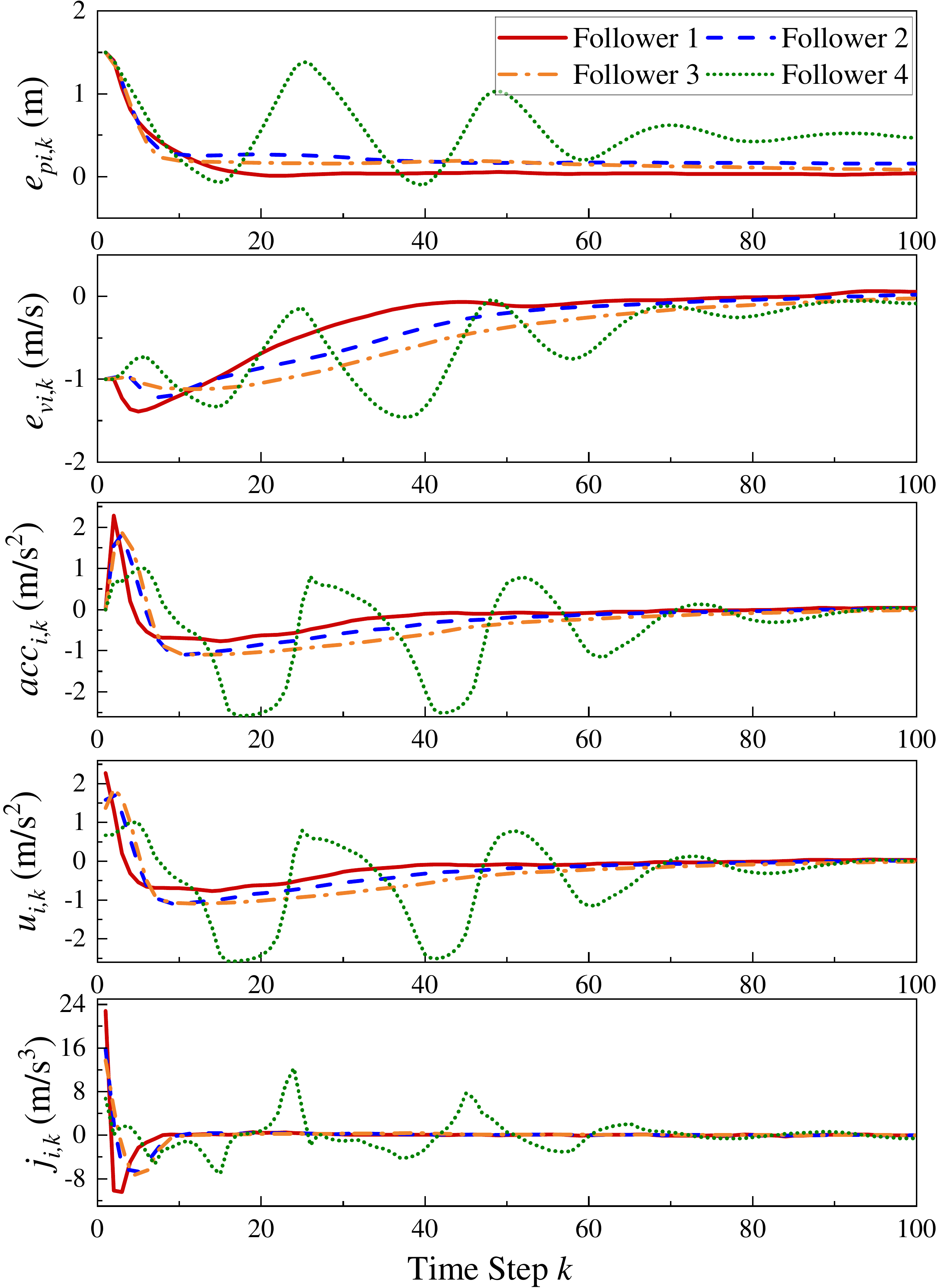}
		\label{fig3_sub1}
	}	
	\subfigure[FH-DDPG]{
            \centering
			\includegraphics[scale=0.3]{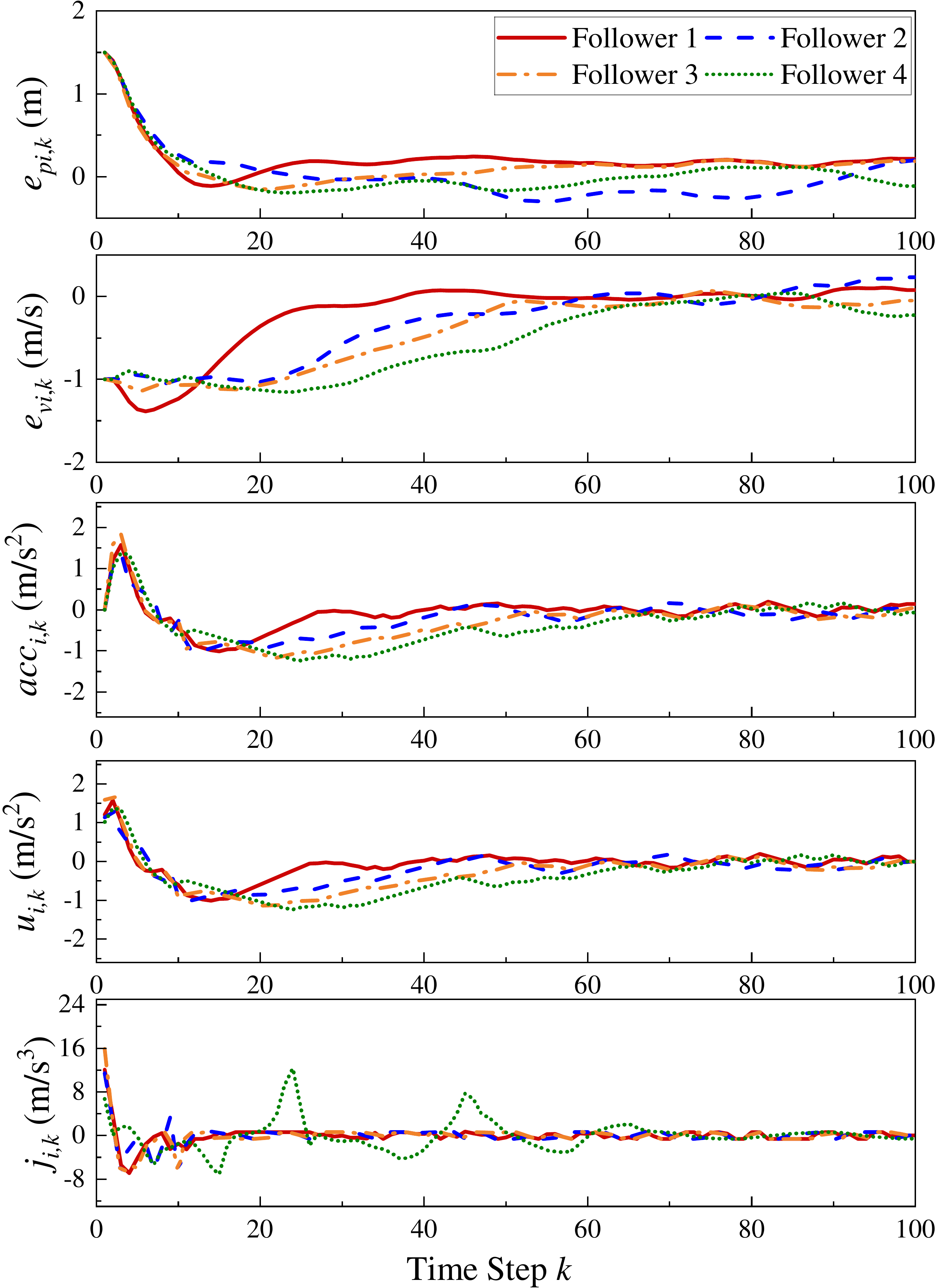}
		\label{fig3_sub2}
	} \par
	\subfigure[HCFS]{
			\includegraphics[scale=0.3]{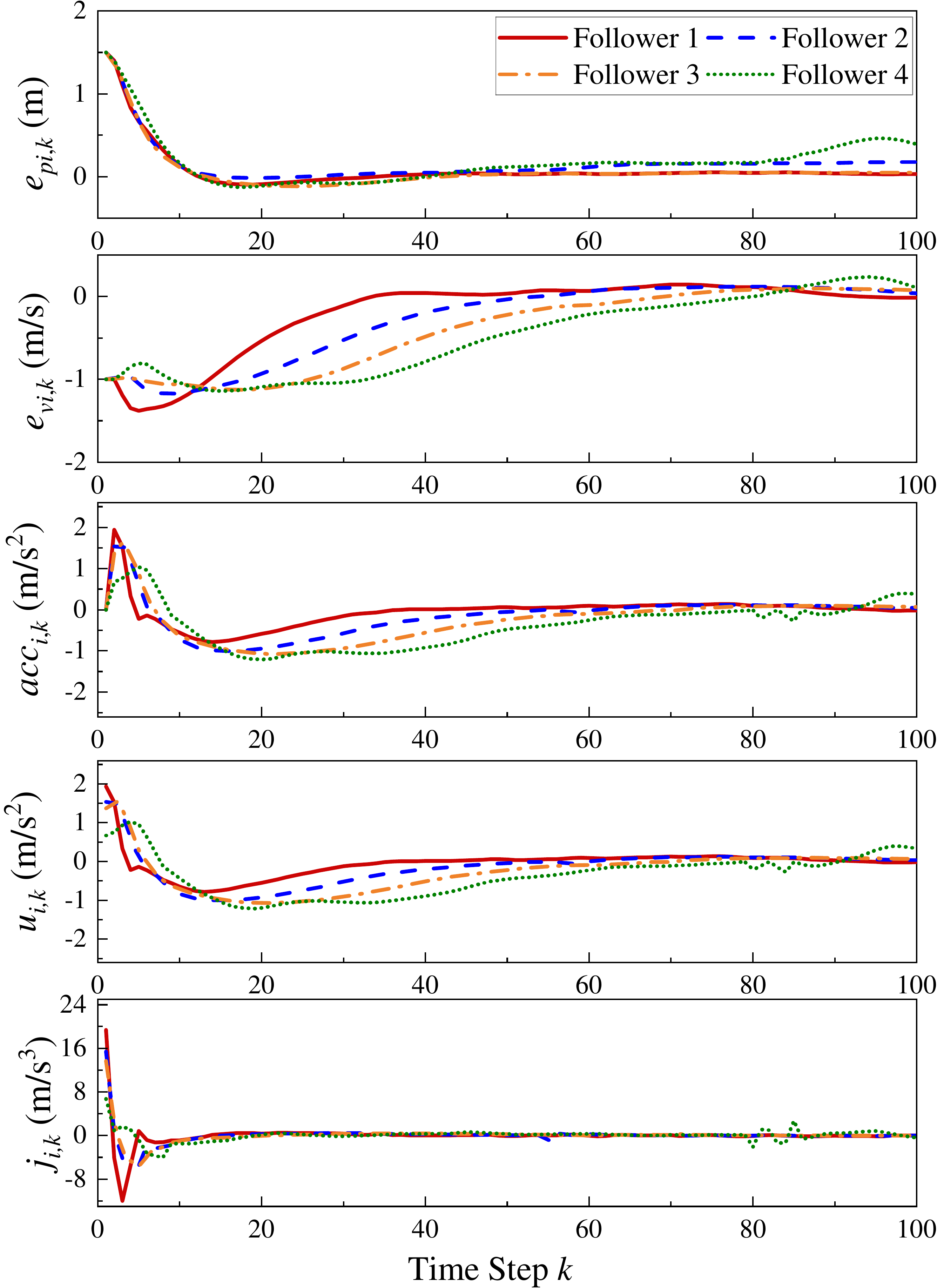}
		\label{fig3_sub1_2}
	}
	\subfigure[FH-DDPG-SS]{
			\includegraphics[scale=0.3]{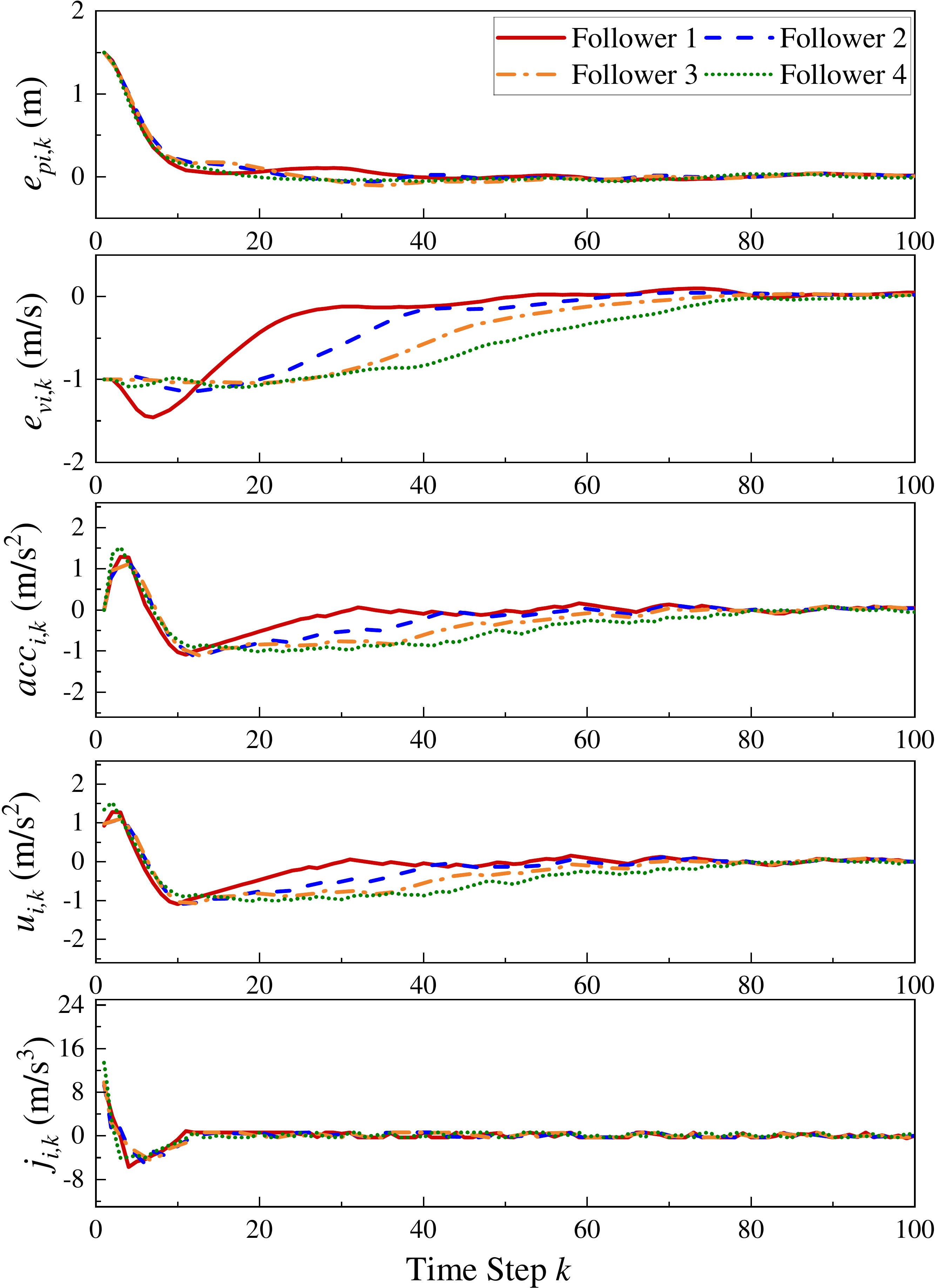}
		\label{fig3_sub3}
	}  
	\caption{Results of a specific test episode. The driving status $e_{pi,k}$, $e_{vi,k}$, and $acc_{i,k}$ along with the control input $u_{i,k}$ and jerk $j_{i,k}$ of each follower $i$ are represented as different curves, respectively.}
	\label{fig_3}
\end{figure*}

	We focus our attention on a specific test episode having $100$ time steps, and plot driving status $e_{pi,k}$, $e_{vi,k}$, $acc_{i,k}$ and control input $u_{i,k}$ along with jerk $j_{i,k}$ of each follower $i$ for all the time steps $k\in\{1,2,\cdots,100\}$. Fig.~\ref{fig3_sub1}, Fig.~\ref{fig3_sub2}, Fig.~\ref{fig3_sub1_2}, and Fig.~\ref{fig3_sub3} show the results of a specific test episode for DDPG, FH-DDPG, HCFS, and FH-DDPG-SS, respectively. It can be observed that the overall shapes of the corresponding curves of all the algorithms look very similar except that the performance curves for follower $4$ using DDPG have large oscillations. This observation is aligned with the results in Table III, where follower $4$ has significantly worse performance when using DDPG compared with using other algorithms. Fig.~\ref{fig_3} shows that in general for each follower $i\in\{1,2,3,4\}$, $e_{pi,k}$ has an initial value of $1.5$ $\rm m$ and is reduced over time to approximately $0$ $\rm m$; $e_{vi,k}$ has an initial value of $-1$ $\rm{m/s}$ and is increased to approximately $0$ $\rm{m/s}$; $u_{i,k}$ is relatively large at the beginning of the episode to increase $acc_{i,k}$ as fast as possible, so that $e_{pi,k}$ and $e_{vi,k}$ can promptly converge to approximately $0$. Correspondingly, $acc_{i,k}$ of each follower $i$ has an initial value of $0$ $\rm m/s^2$ and is suddenly increased to a relatively large value. Then both $u_{i,k}$ and $acc_{i,k}$ are quickly reduced to a negative value, and finally are increased over time to approximately $0$ $\rm {m/s^2}$. After the driving status and control input converge to near $0$, the values fluctuate around $0$ with $u_{i,k}$ trying to maximize the expected cumulative reward in \eqref{eq19} without knowing the future control inputs $u_{i-1,k'}$, $k<k'<K$, of the predecessor $i-1$. Additionally, $j_{i,k}$ of each follower $i$ starts with a large positive value and is then reduced to a negative value. After converging to near $0$ $\rm {m/s^3}$, the value of $j_{i,k}$ fluctuates around $0$ $\rm {m/s^3}$.  

	A closer examination of Fig.~\ref{fig_3} reveals that the performance differences of the algorithms are reflected in convergence speed to steady-state and the oscillations of the driving status and control input. Focusing on $e_{pi,k}$, it can be observed that there are still positive gap-keeping errors for followers $2$, $3$, and $4$ in DDPG up to the end of the episode. Moreover, $e_{pi,k}$ of follower $4$ in DDPG has the slowest convergence speed to $0$ $\rm m$ and the largest oscillations among all the algorithms. Meanwhile, $e_{pi,k}$ in FH-DDPG is reduced to $0$ $\rm m$ for each follower, but there are relatively large oscillations after convergence. $e_{pi,k}$ in HCFS has smaller oscillations than that in FH-DDPG and also converges to $0$ $\rm m$, but there are also large oscillations near the end of the episode for follower $4$. $e_{pi,k}$ in FH-DDPG-SS has the fastest convergence speed to $0$ $\rm m$, and then remains around $0$ $\rm m$ with small oscillations. Now focusing on $e_{vi,k}$, the velocity error in DDPG has the slowest convergence speed to $0$ $\rm m/s$ among all the algorithms. Moreover, $e_{vi,k}$ of follower $4$ in DDPG has the largest oscillations among all the algorithms. $e_{vi,k}$ in FH-DDPG and HCFS has relatively large oscillations after convergence to $0$ $\rm m/s$. $e_{vi,k}$ in FH-DDPG-SS has the smallest oscillations around $0$ $\rm m/s$ after convergence to steady-state. Finally, compared with the other algorithms, FH-DDPG-SS has the smallest jerk $j_{i,k}$ at the beginning of the episode. Although $j_{i,k}$ in FH-DDPG-SS is not as small as those in DDPG and HCFS in the later stage of the episode, it is smaller than that in FH-DDPG and remains at a relatively small level, which can ensure the driving comfort.

\subsection{Platoon safety}	
\begin{figure}[htpb!]
	\centering
	\includegraphics[scale=0.35]{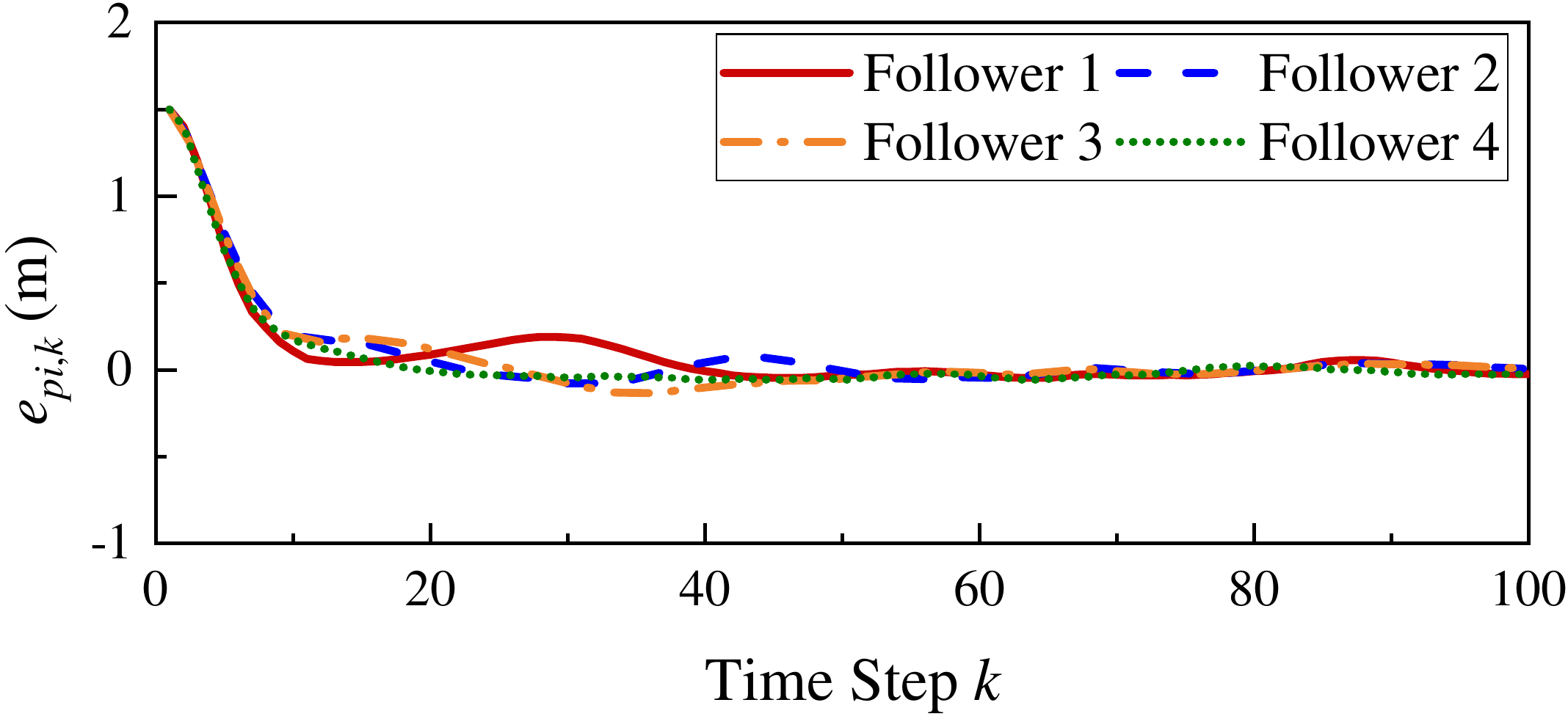}
	\caption{ $e_{pi,k}$ of the worst test episode for FH-DDPG-SS. }
	\label{fig_4}
\end{figure}

	In order to demonstrate that the platoon safety is ensured in the proposed FH-DDPG-SS algorithm, Table \ref{table_5} summarizes the average, maximum, and minimum returns as well as the standard deviation across the 200 test episodes for each follower $i$ in FH-DDPG-SS. Additionally, $e_{pi,k}$ per time step $k$ for the worst test episode among the $200$ test episodes is plotted in Fig. \ref{fig_4}.

	It can be observed from Table \ref{table_5} that the standard deviation of each follower $i$ is small ranging from $0.0015$ to $0.0017$. Moreover, the differences between the maximum and minimum returns are small for all followers. Specifically, the minimum return is worse than the maximum return by $16.28\%$, $15.22\%$, $13.95\%$ and $12.96\%$ for the $4$ followers, respectively.  \par  

\begin{table}[!h]
	\centering
	\setlength\tabcolsep{2.5pt}  
	\caption{The average, maximum, and minimum return as well as the standard deviation across the 200 test episodes for each follower $i$ of FH-DDPG-SS}
	\begin{tabular}{|c|cccc|}
		\hline
		\textbf{Return} & \textbf{Follower 1}& \textbf{Follower 2} &\textbf{Follower 3}&\textbf{Follower 4}\\
		\hline
		Average&-0.0600&-0.0691&-0.0776&-0.0835 \\
		\hline
		Maximum&-0.0559&-0.0644&-0.0731&-0.0787\\
		\hline
		Minimum&-0.0650&-0.0742&-0.0833&-0.0889\\
		\hline
		Standard deviation&0.0015&0.0016&0.0017&0.0017\\
		\hline
	\end{tabular}
	\label{table_5}		
\end{table}

	To demonstrate that platoon safety is ensured even in the worst test episode, it can be observed in Fig. \ref{fig_4} that the followers have an initial gap-keeping error $e_{pi,0}$ of $1.5$ $\rm m$ and the gap-keeping error is reduced over time to approximately $0$ $\rm m$. The most negative $e_{pi,k}$ is $-0.1014$ $\rm m$ at $k=35$, which will not result in vehicle collision since the absolute value of the position error is much smaller than the desired headway.\par  

\subsection{String stability}	

The string stability of a platoon indicates whether oscillations are amplified upstream the traffic flow. The platoon is called string stable if sudden changes in the velocity of a preceding vehicle are attenuated by following vehicles upstream the platoon \cite{Naus2010}. To show the string stability of the proposed FH-DDPG-SS algorithm, we simulate the platoon where the leader acceleration is set to $2$ $\rm{m/s^2}$ when $20<k\leq30$, and $0$ $\rm{m/s^2}$ otherwise. The followers' initial gap-keeping and velocity errors are all set to $0$.\par

\begin{figure}[h!]
	\centering
	\includegraphics[scale=0.35]{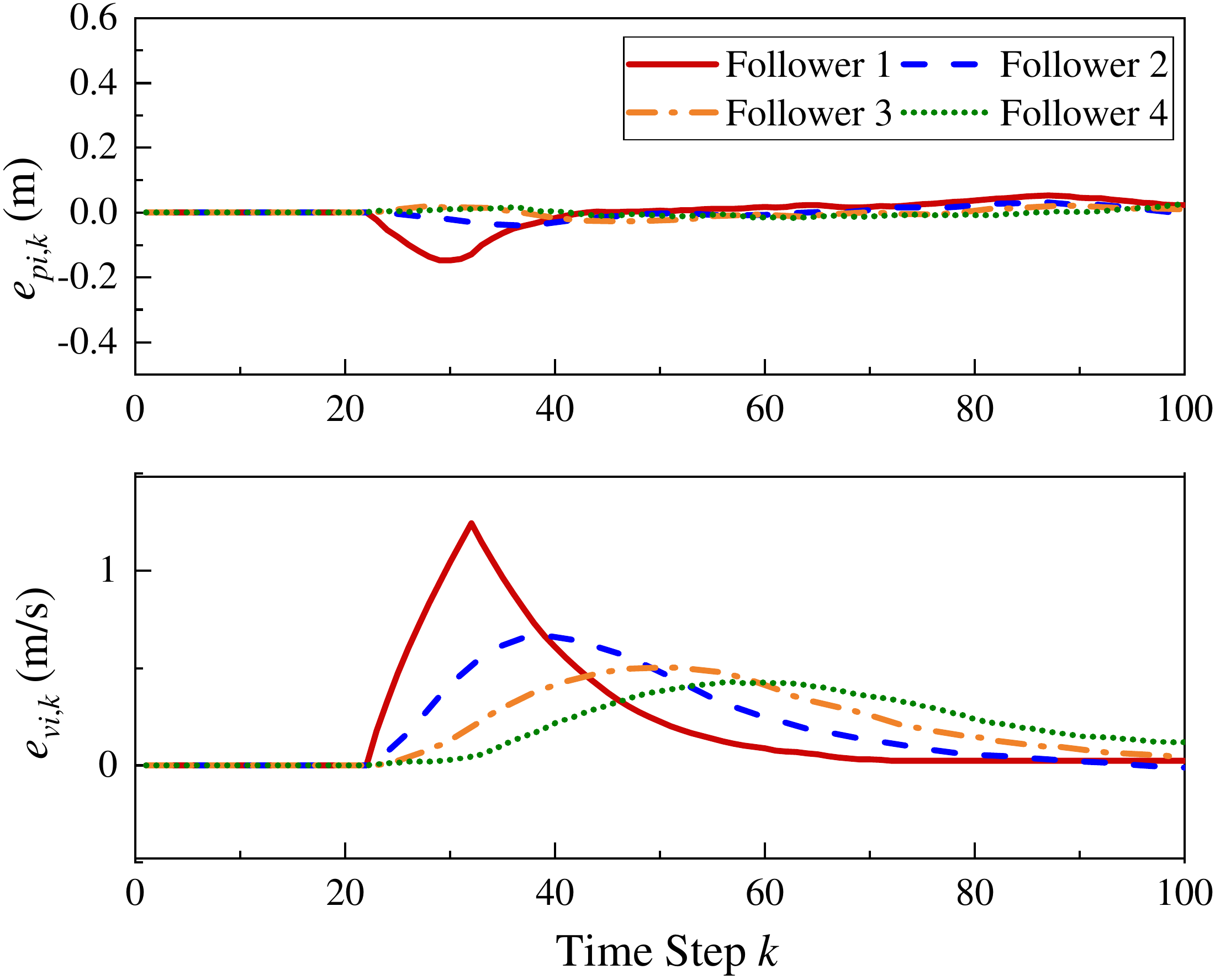}
	\caption{Results of a test episode for FH-DDPG-SS in a specific setting.}
	\label{fig_6}
\end{figure}
As shown in Fig.\ref{fig_6}, the amplitude of the oscillations in $e_{pi,k}$ and $e_{vi,k}$ of each follower $i\in\{2,3,4\}$ are both smaller than those of its respective predecessor $i-1$, demonstrating the string stability of the platoon.

\section{Conclusion}
This paper has studied how to solve the platoon control problem using an integrated DRL and DP method. Firstly, the SSDP model for platoon control has been formulated with a Huber loss function for the reward. Then, the FH-DDPG-SS algorithm has been proposed to improve sampling and training efficiency over the baseline FH-DDPG algorithm with three key ideas. Finally, the performance of FH-DDPG-SS has been compared with DDPG, FH-DDPG and HCFS based on real driving data extracted from the NGSIM. The results have shown that FH-DDPG-SS has learned better policies for platoon control, with significantly improved performance and better convergence stability. Moreover, the platoon safety and string stability for FH-DDPG-SS have been demonstrated.

The focus of this paper is a decentralized platoon control algorithm that optimizes each vehicle's local performance independently. Our future work involves extending the proposed algorithm to deal with multi-agent SSDP, where the objective is to optimize the global performance (i.e., sum of local performances). In addition, we will take into account the actuator delay and V2X communication delay by formulating a random delay SSDP, and develop DRL-based platoon control algorithms with improved performance in the delayed environment.

\appendix

\floatname{algorithm}{Function} 
\setcounter{algorithm}{0}
\subsection{Pseudocode of FH-DDPG Algorithm}
The pseudocode of FH-DDPG algorithm \cite{Lei2020} is given below as Function $\mathrm{FH-DDPG}$.
	\begin{algorithm}
		\renewcommand{\algorithmicrequire}{\textbf{Input:}}
		\renewcommand{\algorithmicensure}{\textbf{Output:}}
		
	\caption{$\mathrm{FH-DDPG}$}
	\label{function1}
	\begin{minipage}{\columnwidth}
	\begin{algorithmic}[1]
	\STATE Randomly initialize actor and critic network weights as $\theta^{\mu0}$ and $\theta^{Q0}$
	\STATE Set $\mu_{i,K}^*(S_{i,K})=\mu^{\mathrm{mo}}(S_{i,K})$ for the final time step $K$
		\FOR{$k= K - 1, \cdots, 1$}
		\STATE $\theta^{\mu_i}=\theta^{\mu0}$ and $\theta^{Q_i}=\theta^{Q0}$ 
		\STATE 	$\theta^{\mu_{i,k}},\theta^{Q_{i,k}}   \leftarrow \mathrm{DDPG-FT}(\theta^{\mu_{i}},\theta^{Q_{i}},\theta^{\mu'_i},\theta^{Q'_i},k)$
		
		\STATE Update the target network:
		\begin{displaymath}
		 \theta^{\mu'_i}\leftarrow\theta^{\mu_{i,k}},\ \theta^{Q'_i}\leftarrow\theta^{Q_{i,k}}
		\end{displaymath}
		\ENDFOR
		\STATE \textbf{return} $\{\theta^{\mu_{i,k}},\theta^{Q_{i,k}}\}_{k=1}^{K} $
		
	\end{algorithmic}
	\end{minipage}
\end{algorithm}

\begin{algorithm}
	\renewcommand{\algorithmicrequire}{\textbf{Input:}}
	\renewcommand{\algorithmicensure}{\textbf{Output:}}
	\caption{$\mathrm{DDPG-FT}(\theta^{\mu_{i}},\theta^{Q_{i}},\theta^{\mu'_i},\theta^{Q'_i},k)$}
	\label{function2}
	\begin{minipage}{\columnwidth}
	\begin{algorithmic}[1]	
		\STATE{Initialize replay buffer $R$}
		\STATE{Initialize a random process $\mathcal{N}$ for action exploration}
		\FOR{episode $e = 1,\dots, E$ }
		\STATE{Receive state $S_{ i,k}^{(e)}$ }
		\STATE{Select action $u_{i,k}^{(e)}$ according to the current policy and exploration noise}
		\STATE{Execute action $u_{i,k}^{(e)}$ and observe reward $r_{i,k}^{(e)}$ and observe new state $S_{i,k+1}^{(e)}$}
		\STATE{Store transition $(S_{i,k}^{(e)},u_{i,k}^{(e)},r_{i,k}^{(e)},S_{i,k+1}^{(e)})$ in $R$}
		\STATE{Sample a random minibatch of $N_b$ transitions $(S_{i,k}^{(n)},u_{i,k}^{(n)},r_{i,k}^{(n)},S_{i,k+1}^{(n)})$ from $R$}
		\IF{$k = K-1$}
		\STATE{Set $y_{i,k}^{(n)}=r_{i,k}^{(n)}+\gamma r_{K} (S_{i,k + 1}^{(n)},\mu^{\mathrm{mo}} (S_{i,k + 1}^{(n)}))$}
		\ELSE
		\STATE{Set $y_{i,k}^{(n)}=r_{i,k}^{(n)}+\gamma Q'_i(S_{i,k + 1}^{(n)},\mu'_i(S_{i,k + 1}^{(n)}|\theta^{\mu'_i})|\theta^{Q'_i})$}
		\ENDIF
		\STATE Update critic by minimizing the loss: 
		\begin{displaymath}
			L=\frac{1}{N_b}\sum_{i} (y_{i,k}^{(n)}-Q_i(S_{i,k}^{(n)},u_{i,k}^{(n)} |\theta^{Q_i}))
		\end{displaymath}
		
		\begin{displaymath}
			\theta^{Q_i}\leftarrow\theta^{Q_i}+\beta \bigtriangledown_{\theta^{Q_i}}L
		\end{displaymath}
		
		\STATE Update the actor using the sampled policy gradient:
		\begin{align}
			 \bigtriangledown_{\theta^{\mu_i}}J\approx&\frac{1}{N_b}(\sum_{i}\bigtriangledown_{u}Q_i(s,u|\theta^{Q_i})|_{s=S_{i,k}^{(n)},u=\mu(S_{i,k}^{(n)})}\IEEEnonumber \\
			&\bigtriangledown_{\theta^{\mu_i}}\mu_i(s|\theta^{\mu_i})|_{S_{i,k}^{(n)}}) \IEEEnonumber
		\end{align}
		
		\begin{displaymath}
			\theta^{\mu_i}\leftarrow\theta^{\mu_i}+\alpha \bigtriangledown_{\theta^{\mu_i}}J
		\end{displaymath}
		
		\ENDFOR

		\STATE \textbf{return} $\theta^{\mu_i}$, $\theta^{Q_i}$
	
		%		\STATE Reset weight of actor and critic networks to initial value:
		%		\begin{displaymath}
			%			\theta^{Q}\leftarrow\theta^{Q0}, \  \theta^{\mu}\leftarrow\theta^{\mu0}
			%		\end{displaymath}
	\end{algorithmic}
\end{minipage}
\end{algorithm}

\bibliographystyle{IEEEtran}
\bibliography{platoon}

\end{document}